\documentclass[fleqn,usenatbib,usedcolumn]{mnras}
\usepackage[british]{babel}             % British English hyphenation
\usepackage{newtxtext}                  % Good fonts
\usepackage[slantedGreek]{newtxmath}    %   "    "   (slanted Greek)
\usepackage[T1]{fontenc}                % Good font encoding
\usepackage{graphicx}                   % Including figures
\usepackage{rotating}
\usepackage{dcolumn}
\newcolumntype{d}{D{.}{.}{-1}}

\def\twolines#1#2{$\kern-6pt\Big\{ {\textrm{#1\hfill}\atop\textrm{#2\hfill}}$}

\usepackage{multirow}
\usepackage{hyperref}
\graphicspath{{.}{./FIGS/}{..}}

\hypersetup{pdfauthor={I. H. Whittam, M. Prescott, K. McAlpine, M. J. Jarvis, I. Heywood},
               pdftitle={The Stripe~82 1-2 GHz Very Large Array Snapshot Survey: host galaxy properties and accretion rates of radio galaxies},
               pdfkeywords={galaxies: active, radio continuum: galaxies, catalogues, surveys},
               bookmarksnumbered=true}
\hypersetup{colorlinks=true,
            linkcolor=blue,
            citecolor=blue,
            filecolor=blue,
            urlcolor=blue}
\volume{{\rm in prep.}}
\title[Radio galaxies in the Stripe 82 VLA survey]{The Stripe~82 1-2 GHz Very Large Array Snapshot Survey: host galaxy properties and accretion rates of radio galaxies}

\author[I.~H.~Whittam et al.]{I.~H.~Whittam$^{1}$\thanks{iwhittam@uwc.ac.za}, M. Prescott$^{1}$, K. McAlpine$^{2}$, M.~J.~Jarvis$^{1,3}$ and I.~Heywood$^{3,4}$\\
   $^{1}$Department of Physics and Astronomy, University of the Western Cape, Robert Sobukwe Road, Bellville 7535, South Africa\\
   $^{2}$SKA South Africa, 3rd Floor, The Park, Park Road, Pinelands, 7405, South Africa\\
   $^{3}$Astrophysics, University of Oxford, Denys Wilkinson Building, Keble Road, Oxford, OX1 3RH, UK\\
   $^{4}$Department of Physics and Electronics, Rhodes University, PO Box 94, Grahamstown, 6140, South Africa\\}

\date{Accepted ---; received ---; in original form ---}

\pagerange{\pageref{firstpage}--\pageref{lastpage}}

\pubyear{2018}
\begin{document}

\label{firstpage}

\maketitle

%@arxiver{accretion2-crop.pdf} 

\begin{abstract}
A sample of 1161 radio galaxies with $0.01 < z < 0.7$ and $10^{21} < L_{1.4~\rm GHz} / \textrm{W Hz}^{-1} < 10^{27}$ is selected from the Stripe 82 1-2 GHz Karl G. Jansky Very Large Array Snapshot Survey, which covers 100 square degrees and has a 1$\sigma$ noise level of 88~$\muup$Jy beam$^{-1}$. Optical spectra are used to classify these sources as high excitation and low excitation radio galaxies (HERGs and LERGs), resulting in 60 HERGs, 149 LERGs and 600 `probable LERGs'. The host galaxies of the LERGs have older stellar populations than those of the HERGs, in agreement with previous results in the literature. We find that the HERGs tend to have higher Eddington-scaled accretion rates than the LERGs but that there is some overlap between the two distributions. We show that the properties of the host galaxies vary continuously with accretion rate, with the most slowly accreting sources having the oldest stellar populations, consistent with the idea that these sources lack a supply of cold gas. We find that 84 per cent of our sample release more than 10 per cent of their accretion power in their jets, showing that mechanical AGN feedback is significantly underestimated in many hydrodynamical simulations. There is a scatter of $\sim 2$~dex in the fraction of the accreted AGN power deposited back into the ISM in mechanical form, showing that the assumption in many simulations that there is a direct scaling between accretion rate and radio-mode feedback does not necessarily hold.
We also find that mechanical feedback is significant for many of the HERGs in our sample as well as the LERGs.
\end{abstract}


%------------------------------------------------------------------------------%
\section{Introduction}\label{section:intro}

Essentially all bulge dominated galaxies have a supermassive black hole at or near their centre (e.g. \citealt{1998AJ....115.2285M}) and in some galaxies, accretion of matter onto this central black hole powers an Active Galactic Nucleus (AGN). Depending on the selection method, approximately 10 - 20 per cent of AGN are radio loud, often displaying powerful radio jets which are many hundreds to thousands of times the size of the host galaxy. 

Over the years several ways of classifying AGN have emerged, based on observations in different parts of the electromagnetic spectrum (e.g.\ the Seyfert 1 and 2 classifications based on optical spectra and the Fanaroff and Riley I and II categories based on radio morphology \citealt{1974MNRAS.167P..31F}), see \citet{2017A&ARv..25....2P} for a recent review. While the differences between several classes of AGN can be explained by orientation effects \citep{1993ARA&A..31..473A,1995PASP..107..803U}, the differing properties of some AGN cannot be unified in this way. A significant number of radio loud sources lack the high-excitation lines in their optical spectra which are characteristic of `traditional' AGN \citep{1979MNRAS.188..111H,1994ASPC...54..201L}, and many seem to lack a dusty torus (e.g.\ \citealt{2004ApJ...602..116W,2006ApJ...647..161O}). Several also show no evidence of accretion related X-ray emission \citep{2006MNRAS.370.1893H}. These differences cannot be explained by orientation effects.

This resulted in the idea that there are two fundamentally different classes of radio AGN; one class which posses the `traditional' accretion related structures such as an accretion disc and a dusty torus (referred to as High Excitation Radio Galaxies, HERGs, due to the high-excitation lines visible in their optical spectra), and a second class of radio-loud AGN which are missing these structures (Low Excitation Radio Galaxies, or LERGs). HERGs radiate efficiently across the whole electromagnetic spectrum, while LERGs radiate inefficiently and emit most of their energy in kinetic form as powerful jets (e.g.\ \citealt{2007MNRAS.381..589M,2009MNRAS.396.1929H}). HERGs also seem to be more dominated by emission from their cores than LERGs \citep{2016MNRAS.462.2122W}. However, \citet{2014MNRAS.440..269M} found seven sources which are classified as LERGs based on their optical spectra but which show signs of being radiatively efficient at other wavelengths, suggesting that the picture is not always so clear cut. As HERGs are thought to have tori due to the prevalence of hot dust emission (e.g.\ \citealt{2014MNRAS.438.1149G}), their observed properties are expected to be orientation dependent, while the properties of LERGs are not (e.g.\ \citealt{2011MNRAS.417.2651M}). 

The current understanding is that these two classes are a result of two fundamentally distinct accretion modes \citep{2005MNRAS.362...25B,2006MNRAS.370.1893H}. HERGs are thought to accrete cold gas efficiently via an optically thick, geometrically thin accretion disc (e.g.\ \citealt{1973A&A....24..337S}), while LERGs accrete from a hot gas reservoir (e.g.\ \citealt{2012A&A...541A..62J,2014ARA&A..52..529Y,2007MNRAS.376.1849H}) relatively slowly via an advection-dominated accretion flow \citep{1995ApJ...452..710N,2003ANS...324..435Q}. As a result, the two classes are sometimes referred to as `cold mode' and `hot mode' sources respectively (the terms `radiative mode' and `jet mode' are also sometimes used). These two accretion modes appear to be strongly influenced by environment (e.g.\ \citealt{2008A&A...490..893T,2017MNRAS.466.4346M}).

Evidence suggests that the two classes reside in host galaxies with differing properties; HERGs tend to be found in less massive galaxies with a young stellar population, while LERGs are hosted by massive galaxies, often at the centre of a group or cluster \citep{1990AJ.....99...14B,2007MNRAS.379..894B}, which mostly contain older stars (e.g.\ \citealt{2009ApJ...699L..43S,2010MNRAS.406.1841H,2012MNRAS.421.1569B,2017MNRAS.466.4346M,2017MNRAS.464.1306C}). \citet{2017ApJ...847...72B} found that LERGs at $z \sim 1$ have been devoid of significant star-formation activity for longer than 1 Gyr. This fits in with the favoured scenario where HERGs have a large supply of cold gas which they accrete efficiently, as this will also provide the fuel for star formation, leading to a young stellar population. LERGs, on the other hand, lack this cold gas supply so have a low star-formation rate, and instead accrete less efficiently from the warmer gas that they have available. HERGs also show rapid redshift evolution, while LERGs show little or none out to $z\sim1$ \citep{2004MNRAS.352..909C,2014MNRAS.445..955B,2016MNRAS.460....2P,2016MNRAS.457..730P}. Recent work by \citet{2018MNRAS.tmp..105W} has found that at higher redshifts (out to $z \sim 2$) LERGs show strong negative evolution.

The link between star formation and AGN activity in a galaxy has been widely studied in the last decade and is not yet well understood (see reviews by \citealt{2009Natur.460..213C,2012ARA&A..50..455F,2014ARA&A..52..589H}). It is suggested that there is a link between the accretion of matter onto the central black hole and the star formation activity that occurs in a galaxy, via the gas supply available. Models of galaxy evolution require star formation activity to be quenched, and in more massive galaxies it is widely thought that AGN activity may be responsible for this (e.g.\ \citealt{2005Natur.433..604D,2006MNRAS.365...11C,2006ApJS..163....1H,2010ApJ...717..708C}). However, there is recent theoretical \citep{2013ApJ...772..112S} and observational \citep{2014MNRAS.442.1181K} evidence that AGN feedback can in fact have a positive effect on star formation. It is thought that the different AGN accretion modes have different feedback effects on the host galaxy (see review by \citealt{2012ARA&A..50..455F}), with HERGs associated with `quasar mode' feedback and LERGs with `radio mode'. Understanding AGN accretion is therefore crucial to understanding the history of galaxy evolution across cosmic time. 

In order to build up a complete picture of AGN activity, we need information from across the electromagnetic spectrum. 
In this work we use a sample of $\sim 1000$ radio galaxies selected from a 100~square degree NSF's Karl G. Jansky Very Large Array (VLA) survey at $1 - 2$~GHz in the Sloan Digital Sky Survey (SDSS) Stripe 82 \citep{2016MNRAS.460.4433H}, along with optical spectroscopy and mid-infrared data from \emph{Wide-field Infrared Survey Explorer} (\emph{WISE}; \citealt{2010AJ....140.1868W}). The Stripe 82 radio data used here is a factor of two deeper than the Faint Images of the Radio Sky at Twenty-one cm (FIRST), meaning that we can probe less powerful sources than studies based on FIRST or NVSS, such as \citet{2012MNRAS.421.1569B} and \citet{2017MNRAS.464.1306C}, while the areal coverage of $100~\rm deg^2$ provides enough cosmological volume for statistical studies. The primary aim of this paper is to investigate the relationship between accretion rate onto the central black hole and the properties of the galaxies by combining the radio observations with optical spectroscopy and mid-infrared observations. It is only possible to detect LERGs in the radio, meaning they are generally missing from AGN samples selected at other wavelengths, resulting in an incomplete picture of AGN activity.

This paper is laid out as follows; in Section~\ref{section:data} we describe the data used in this work and the sample selection. The sources are classified as HERGs and LERGs, using the scheme outlined in Section~\ref{section:classifying}, and the properties of these HERGs and LERGs are discussed in Section~\ref{section:properties}. In Section~\ref{section:rates} we present the accretion rates of the different source classifications, and investigate the properties of the galaxies in the sample as a function of accretion rate. The accretion rates of the different classes and the implications for AGN feedback in hydrodynamical simulations are discussed in Section~\ref{section:discussion}. The conclusions are presented in Section~\ref{section:conclusions}. Throughout this paper the following values for the cosmological parameters are used: $H_0 = 70~{\rm km \, s}^{-1} ~\rm Mpc^{-1}$, $\Omegaup_{\rm M} = 0.3$ and $\Omegaup_{\Lambdaup} = 0.7$.

%------------------------------------------------------------------------------%
\section{Data used}\label{section:data}

\subsection{Radio data}

This work is based on sources selected from a 1-2~GHz survey of SDSS Stripe~82 made with the VLA by \citet{2016MNRAS.460.4433H}. This survey covers $\sim 100$~deg$^2$ and was carried out in the hybrid CnB configuration of the VLA, giving a resolution of $16 \times 10$~arcsec. The survey consists of 1026 snapshots each of 2.5~minutes duration, giving a typical noise in the resulting mosaic of $88~\muup$Jy beam$^{-1}$. The compact configuration used for this survey means that it has excellent sensitivity to diffuse, low-surface brightness structures. 
The final source catalogue contains 11,782 components, which make up approximately 8,948 unique radio sources. For full details of the data reduction, imaging and source finding, we refer the reader to \citet{2016MNRAS.460.4433H}.

A second VLA radio survey at 1.4~GHz, taken by \citet{2011AJ....142....3H}, is also available in the Stripe~82 field. This survey was carried out in A configuration resulting in an angular resolution of 1.8~arcsec and has a mean rms noise of 52~$\muup$Jy beam$^{-1}$. The \citeauthor{2011AJ....142....3H} survey therefore has a higher resolution than the \citeauthor{2016MNRAS.460.4433H}. survey, but resolves out emission on larger scales, and several sources which appear as single point sources in the \citeauthor{2011AJ....142....3H} data are in fact part of larger, multi-component sources when viewed in the \citeauthor{2016MNRAS.460.4433H} data. As many radio AGN are extended structures with diffuse emission in their lobes, the \citeauthor{2016MNRAS.460.4433H} survey provides a better measurement of their total flux density so is used as the starting point for this work. Importantly however, the \citeauthor{2011AJ....142....3H} data provides the positional accuracy required to identify multi-wavelength counterparts to the radio sources.

\subsection{Optical data}\label{section:optical}

The Stripe~82 field was repeatedly observed as part of the SDSS survey \citep{2000AJ....120.1579Y} in $u$, $g$, $r$, $i$, $z$ and thus coadded images can be produced that are $\sim2$ mags fainter than the Main Galaxy Survey, reaching a depth of $g \sim 24.5$ \citep{2014ApJ...794..120A}. Optical spectra are available for a large number of sources in the field from the SDSS I - IV and Baryon Oscillation Spectroscopic Survey (BOSS) observing campaigns. In this work, we use the the value-added spectroscopic catalogues based on these spectra from the SDSS 14th Data Release (SDSS DR14; \citealt{2017arXiv170709322A}) produced by the Portsmouth Group, available at \url{http://www.sdss.org/dr14/spectro/galaxy_portsmouth/}. The procedures used to obtain the emission line fluxes and stellar kinematics are described in \citet{2013MNRAS.431.1383T} and the spectro-photometric model fitted used to estimate stellar masses, star-formation rates and other galaxy properties are explained in \citet{2009MNRAS.394L.107M}. In this work we use the stellar masses determined from passively evolving galaxy templates.

\subsection{Cross-matching}

In order to optimise reliability when identifying optical counterparts for the radio sources the cross-matching was carried out by eye using the python script \textsc{Xmatchit}\footnote{available at \url{https://github.com/MattPrescottAstro/}}, as described in \citet{2018MNRAS..tempP}. For each component in the \citet{2016MNRAS.460.4433H} radio catalogue two cut-outs with different zoom levels were produced, overlaying radio contours from both the \citeauthor{2016MNRAS.460.4433H} and \citeauthor{2011AJ....142....3H} surveys on to the co-added SDSS $r$-band optical images. The positions of the \citeauthor{2016MNRAS.460.4433H} radio components and the SDSS optical objects were also shown in each image, allowing the user to visually identify the most likely optical counterpart for each radio component. Each radio component was classified by two people, and in any cases where there was a discrepancy between the two classifications the source was examined by a committee of three people and a final classification was decided upon.

Out of the $11 768$ radio source components in the \citeauthor{2016MNRAS.460.4433H} catalogue, $6 754$ (57 per cent) have an optical counterpart identified. As many of the radio sources consist of more than one component, the final matched catalogue contains $4 795$ unique sources. $2 799$ of these matches are to objects in the photometric catalogue, and $1 997$ to objects in the spectroscopic catalogue. See \citet{2018MNRAS..tempP} for further details of the matching process and the reliability and completeness of the matched catalogues.

\subsection{Sample used}\label{section:sample_used}

In this work we consider the sample of \citeauthor{2016MNRAS.460.4433H} radio sources which have a counterpart in the spectroscopic catalogue described in Section~\ref{section:optical}. We restrict our analysis to sources with $z < 0.7$, which is a conservative cut to avoid incompleteness, meaning that we have 1501 sources in our sample.

As we require radio sources to have an optical spectrum available to be included in our sample, in the redshift range 0.4 < z < 0.7 we are limited to radio sources selected for observation as part of the BOSS survey. BOSS employs a colour selection, which means that our radio sample may be missing some blue objects at z > 0.4. We do not expect this to affect the results of this paper, this is discussed further in Section~\ref{section:galproperties}.

\subsection{\emph{WISE} data}\label{section:WISE_data}

The \emph{Wide-field Infrared Survey Explorer} (\emph{WISE}; \citealt{2010AJ....140.1868W}) telescope has observed the whole sky in four infrared bands: W1 (3.4~$\muup$m), W2 (4.6~$\muup$m), W3 (12~$\muup$m) and W4 (22~$\muup$m) with angular resolutions of 6.1, 6.4, 6.5 and 12 arcsec respectively. The $5\sigma$ point source sensitivities in the four bands is better than 0.08, 0.11, 1 and 6 mJy respectively in unconfused regions on the ecliptic.

We searched the \emph{WISE} All-Sky catalogue for counterparts to each radio source with an optical counterpart using a search radius of 2~arcsec to the optical position of each source. 1444 out of 1501 radio sources have a match, however, two-thirds of these sources only have a detection in the more sensitive W1 and W2 bands. 

The rest-frame 12-$\muup$m luminosities are k-corrected using the spectral index between 12 and 22~$\muup$m for sources with a detection in both bands, in a similar way to \citet{2014MNRAS.438.1149G}. For sources with an upper limit in one or both bands the mean spectral index for the sample was used instead (mean $\alpha = 2.53 \pm 0.02$, where $S \propto \nu^{-\alpha}$, calculated using sources with a detection in both bands only).

%------------------------------------------------------------------------------%
\section{Classifying the radio galaxies}\label{section:classifying}

\subsection{Identifying the star-forming galaxies}

In order to study the properties of the AGN in this sample, we first need to separate out the star-forming galaxies that are present in the radio catalogue. To do this we use a combination of three different criteria: the D$_{\rm n}$(4000) against $L_{1.4~\rm GHz} / M_{\ast}$ separation developed by \citet{2005MNRAS.362....9B}, the `BPT' diagram \citep{1981PASP...93....5B}, and a cut in the luminosity of the H$\alpha$ line ($L_{H\alpha}$) against $L_{1.4~\rm GHz}$ plane. We refer the reader to \citet{2018MNRAS..tempP} for full details of this classification scheme. Note that these criteria are very similar to those used by \citet{2012MNRAS.421.1569B} to separate star-forming galaxies and AGN.

In total, 340 sources were identified as star-forming, leaving 1161 radio AGN in our sample.

\subsection{Classifying the HERGs and LERGs}\label{section:classifications}

\begin{figure}
\centerline{\includegraphics[width=\columnwidth]{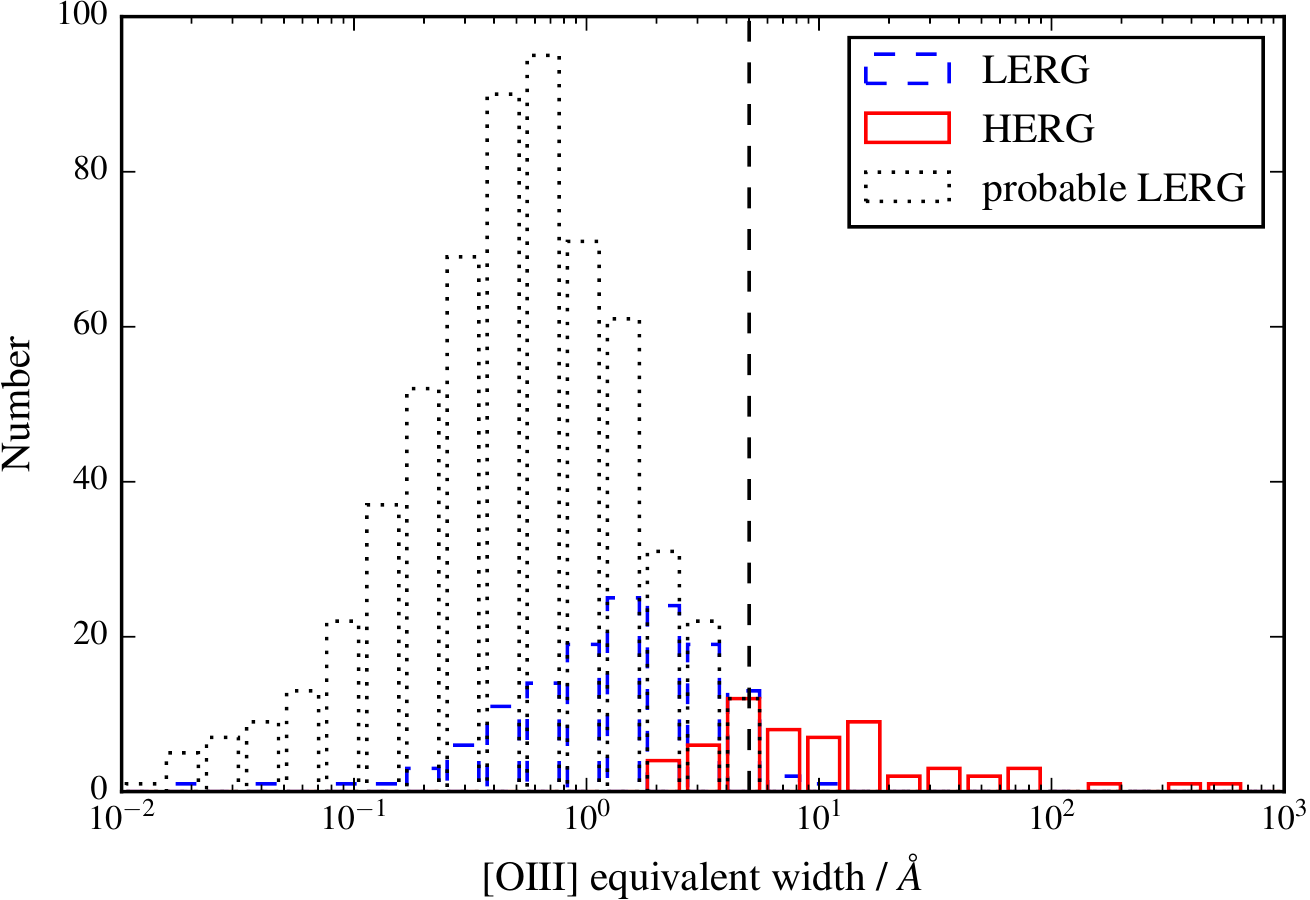}}
\caption{Distribution of [OIII] equivalent widths, showing sources classified as HERGs, LERGs and probable LERGs. The cut off used to classify sources as probable LERGs is shown by the vertical dashed line.}\label{fig:OIIIEW}
\end{figure}

Sources were classified as HERGs or LERGs in the same way as those in \citet{2012MNRAS.421.1569B} to allow a direct comparison with their work. This classification scheme uses a combination of line ratios and [OIII] equivalent width, as described below.  Lines must have an Amplitude-over-Noise (AoN) $>1.5$ to be considered a detection.

1) The Excitation Index ($EI$; \citealt{2010A&A...509A...6B}) was calculated using the following expression: $EI = \rm \textrm{ log}_{10}([OIII]\lambda 5007/H\beta) - \frac{1}{3}[\textrm{ log}_{10}([NII]\lambda 6583 /H\alpha) + \textrm{ log}_{10}([SII]\lambda \lambda 6716,6731/H\alpha) + \textrm{log} ([OI]\lambda 6364 /H\alpha)]$. Sources with all six lines detected and an $EI$ value at least $1\sigma$ greater than 0.95 were classified as HERGs, and those with an $EI$ value at least $1\sigma$ below 0.95 were classified as LERGs.

2) If the four lines required for one of the \citet{2006MNRAS.372..961K} versions of the BPT diagram were detected and a source lies at least $1\sigma$ away from the Seyfert-LINER dividing line, then sources in the Seyfert region were classified as HERGs and sources in the LINER region as LERGs. 

3) If the [OIII] equivalent width was more than $1\sigma$ greater than $5~$\AA, then the source was classified as a HERG.

4) Stages 1) -- 3) were repeated for the unclassified sources, removing the requirement to be at least $1\sigma$ away from the dividing value.

5) The [NII]/H$\alpha$ against [OIII]/H$\alpha$ diagnostic of \citet{2010MNRAS.403.1036C} was used if the four lines needed were detected. If the AoN value for the [OIII] line was $<1.5$ then the flux value was considered an upper limit, which allowed some additional sources to be classified as LERGs.

Sources which we are unable to classify using the criteria listed above but which have an [OIII] equivalent width $<5$~\AA~and a signal to noise ratio of the [OIII] line measurement $>3$ are classified as `probable LERGs'. This allows us to compare more directly with studies such as \citet{2017MNRAS.464.1306C} which only use an [OIII] equivalent width cut to classify HERGs and LERGs. The dividing value of 5~\AA~is chosen as it is widely used in the literature. Fig.~\ref{fig:OIIIEW} shows the distribution of equivalent widths for the different source classes. Note that a small number of HERGs have [OIII] equivalent widths less than 5~\AA, the cut off used to classify sources as `probable LERGs', meaning that it is possible that some of these `probable LERG' sources could be HERGs. However, there is only a very small overlap between the HERG and `probable LERG' distributions, and the vast majority of the `probable LERGs' have [OIII] equivalent widths much less than any of the HERGs meaning they are most likely to be LERGs, hence the name of the class. 
Note that we do not see the bimodal distribution of equivalent widths which was originally used to separate HERGs and LERGs, which we attribute to the lower radio luminosities we are sensitive to.

81 sources are identified as QSOs by the SDSS pipeline so are classified as such. These sources are excluded from the rest of this work. 

271 sources do not have the line measurements required to be classified into any of the categories so remain unclassified. As these sources do not have strong enough lines to be measured, they are much more likely to be LERGs than HERGs.

The total number of sources in each category is as follows; HERGs = 60, LERGs = 149, probable LERGs =  600, QSOs = 81 and unclassified sources = 271, with 340 SFGs. LERGs are therefore the dominant population, accounting for 84 per cent of the classified AGN sample if the `probable LERG' class are included. LERGs make up 61 per cent of the sample if SFGs are included as well, which is slightly lower than \citet{2017MNRAS.464.1306C} who find that 70 per cent of their FIRST and SDSS-selected sample with z < 0.8 are LERGs. This is due to the increased fraction of star-forming galaxies in our deeper radio survey.

%------------------------------------------------------------------------------%
\section{Properties of the radio galaxies}\label{section:properties}

\begin{figure}
\centerline{\includegraphics[width=\columnwidth]{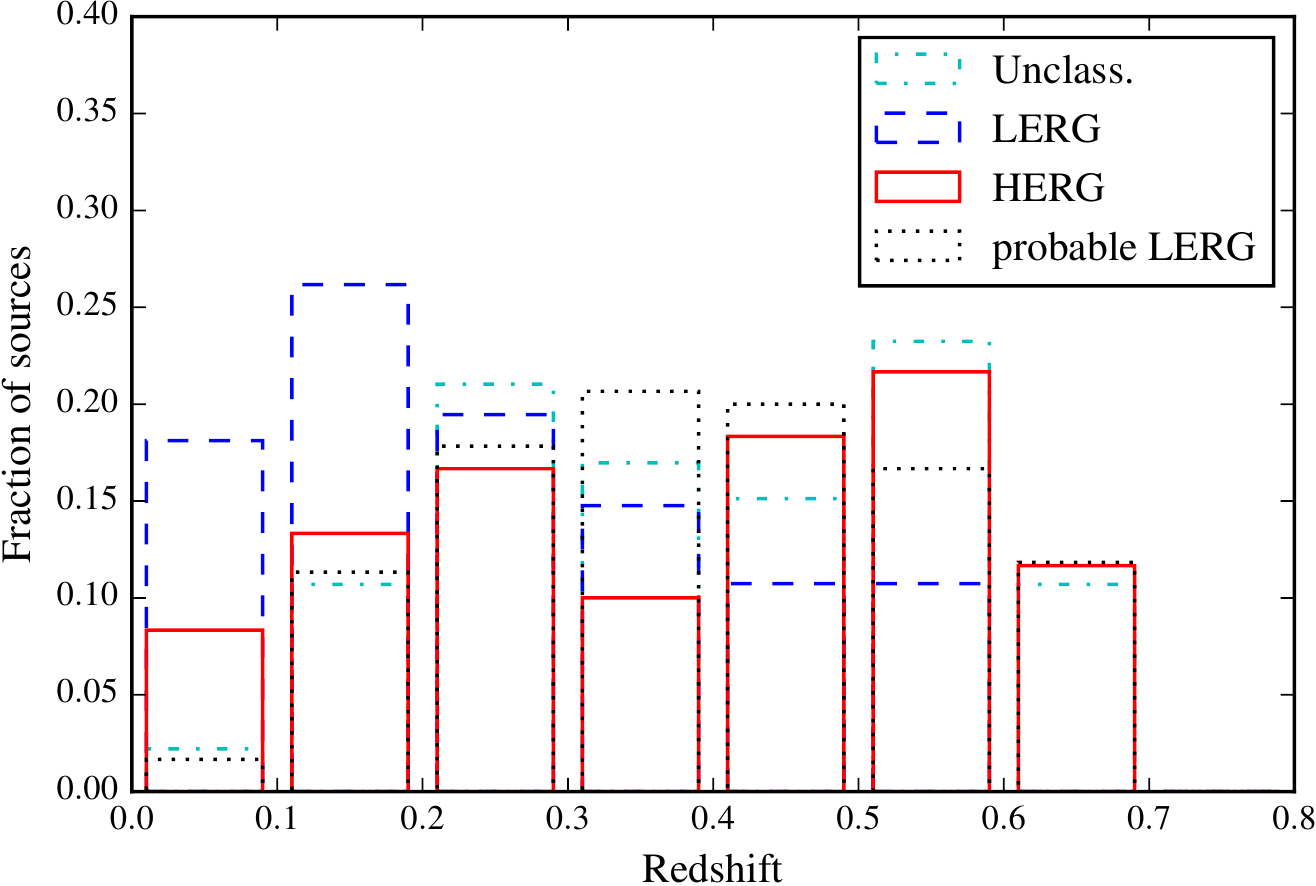}}
\medskip
\centerline{\includegraphics[width=\columnwidth]{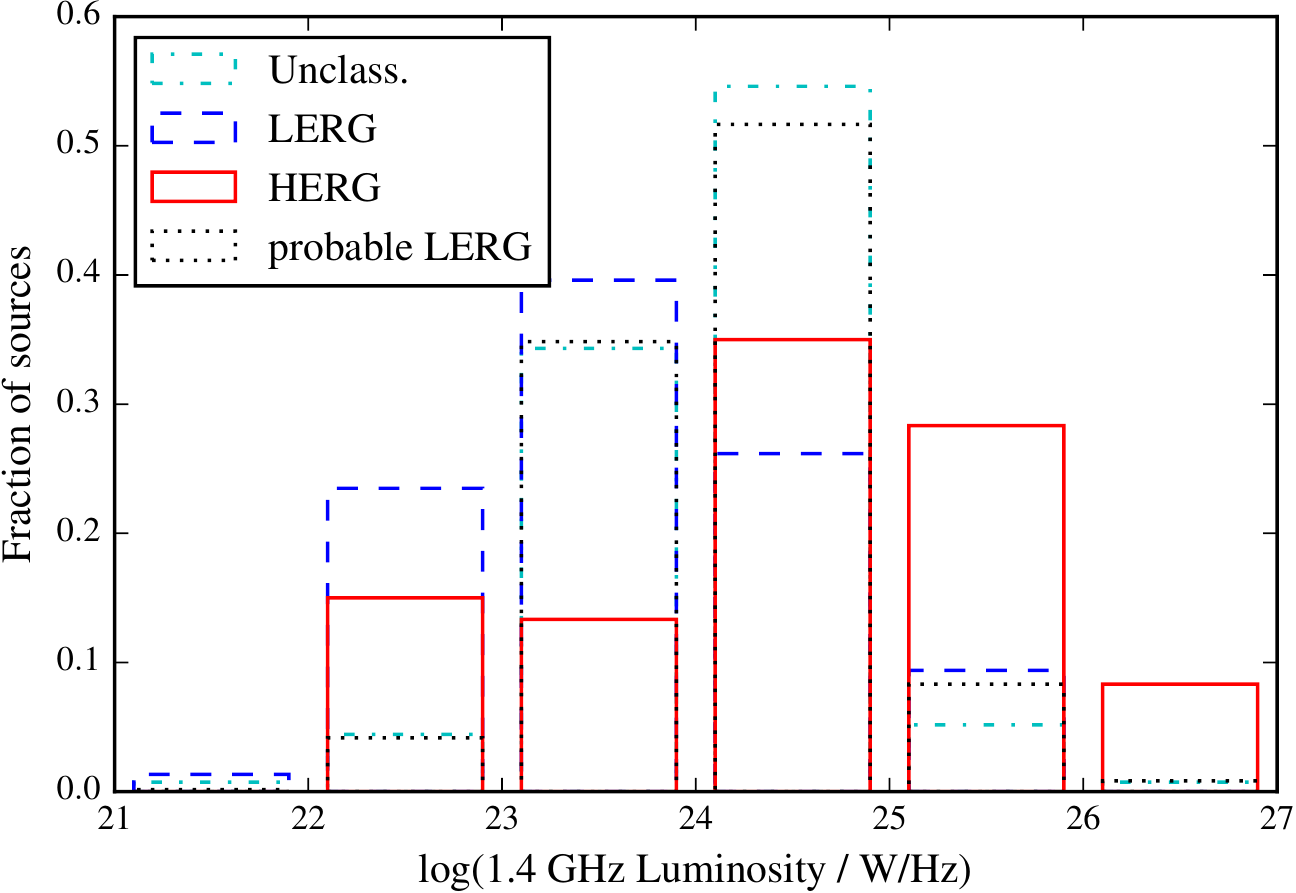}}
\caption{Redshift (top panel) and 1.4~GHz luminosity distribution (bottom panel) of the sources classified as HERGs and LERGs. The distributions are normalised by the total number of sources in each classification.}\label{fig:zlum}
\end{figure}

\begin{figure}
\centerline{\includegraphics[width=\columnwidth]{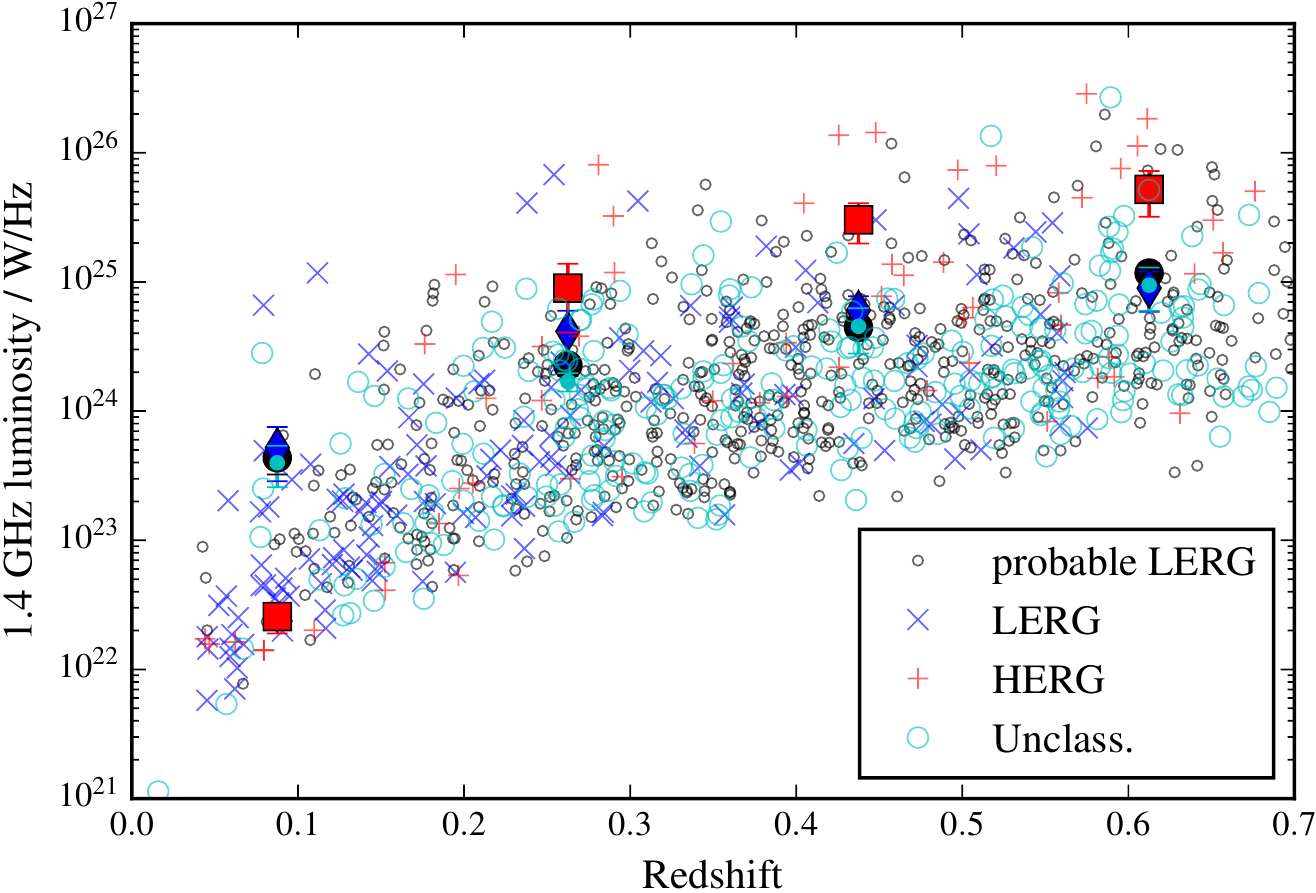}}
\caption{1.4-GHz luminosity as a function of redshift for HERGs, LERGs, probable LERGs and unclassified sources. Note that the method used to classify HERGs and LERGs is different to that used in \citet{2018MNRAS..tempP}, which explains the differences between this plot and the similar one in that paper.}\label{fig:zlum2}
\end{figure}

The redshift and luminosity distributions of the sources classified as HERGs and LERGs are shown in Figs.~\ref{fig:zlum} and \ref{fig:zlum2}. The radio luminosities are k-corrected by assuming a spectral index of 0.7 (the convention $S \propto \nu^{-\alpha}$ is used throughout this work). The HERGs tend to be found at higher redshifts than the LERGs, with a median redshift of 0.40 for the HERGs compared to 0.23 for the LERGs. There is a tendency for the `probable LERGs' to have higher redshifts than the securely classified LERGs (probable LERG median redshift = 0.39), which is because the higher-redshift LERGs are less likely to have enough detected lines to be classified definitively as LERGs using the full \citet{2012MNRAS.421.1569B} criteria.

Although the HERGs tend to have higher radio luminosities than the LERGs, with median luminosities of $2.9 \times 10^{24}~\rm{W \, Hz^{-1}}$ compared $4.3 \times 10^{23}~\rm{W \, Hz^{-1}}$, HERGs are found across the full range of radio luminosities, down to $L_{1.4~\rm GHz} \sim 10^{22}~\rm{W \, Hz^{-1}}$. This agrees with \citet{2012MNRAS.421.1569B}, but goes against assumptions often made in the literature. \citet{2015MNRAS.447.1184F} find that the luminosity distributions of HERGs and LERGs at $z \sim 1$ are indistinguishable up to $L_{151~\rm MHz} =  3 \times 10^{27}~\rm W \, Hz^{-1} \, sr^{-1}$, which is not supported by our results, although our sample is at lower redshift.

The unclassified sources have very similar redshift and luminosity distributions to the probable LERGs, suggesting that they may be drawn from the same population.

\begin{figure}
\centerline{\includegraphics[width=\columnwidth]{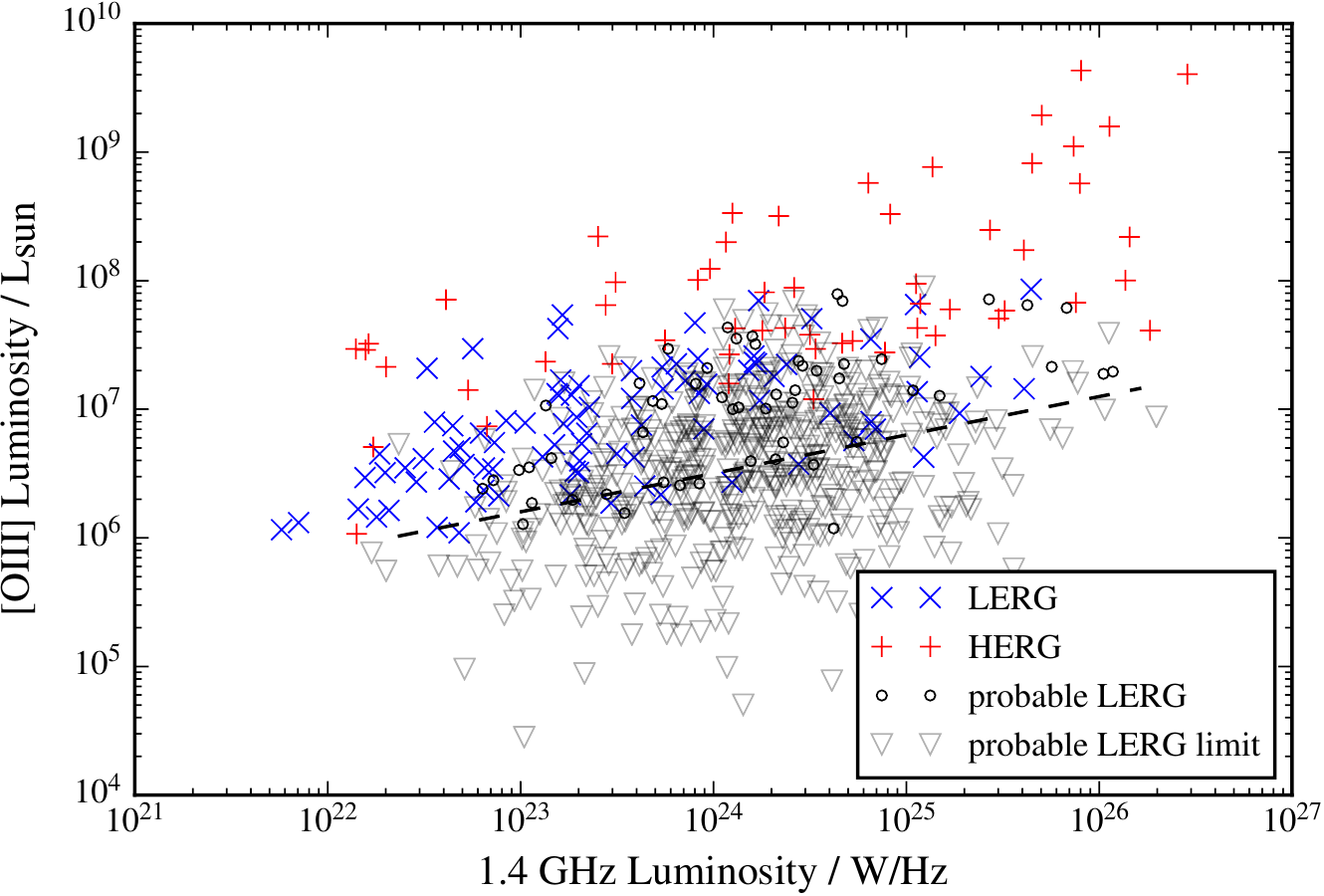}}
\caption{[OIII] line luminosity as a function of 1.4 GHz radio luminosity for the samples defined in Section~\ref{section:classifications}. Triangles show sources with an upper limit on [OIII] luminosity (these are sources with an amplitude-over-noise < 1.5 on the [OIII] line measurement). The dashed line shows the classification proposed by \citet{2012MNRAS.421.1569B}. The unclassified sources do not have an [OIII] line measurement so are unable to be included in this Figure.}\label{fig:LOIIIL14}
\end{figure}

Fig.~\ref{fig:LOIIIL14} shows the [OIII] line luminosity ($L_{\rm [OIII]}$) as a function of 1.4-GHz luminosity ($L_{1.4~\rm GHz}$). For sources with an Amplitude over Noise (AoN) on the [OIII] line detection $<1.5$ the resulting [OIII] luminosity is treated as an upper limit. The relationship between [OIII] and radio luminosity tells us about the relationship between the radiative power (traced by $L_{\rm [OIII]}$) and the jet power (traced by $L_{1.4~\rm GHz}$) of an AGN. There is a clear divide between the HERGs and LERGs in this Figure with the HERGs found above the LERGs, meaning that for a given jet power, HERGs have a higher radiative power. This is consistent with the current understanding where LERGs radiate inefficiently, emitting the bulk of their power via their jets. This result is similar to that found by several studies such as \citet{2009MNRAS.396.1929H,2010A&A...509A...6B,2012MNRAS.421.1569B}, with the latter suggesting that this diagram could be used to classify sources at $z < 0.1$. They propose that all sources below the dashed line plotted in Fig.~\ref{fig:LOIIIL14} can be securely classified as LERGs. This is backed up by our sample as there are no HERGs below this line, but note that many LERGs would be missed by this selection and a similar method cannot be used to select HERGs.

\subsection{Host galaxy properties}\label{section:galproperties}

Information from optical spectra allows us to constrain the properties of the host galaxies by providing information such as stellar age and stellar mass. Understanding these properties is key to building up a complete picture of an AGN and therefore understanding the evolution of AGN and star-formation across cosmic time.

\begin{figure}
\centerline{\includegraphics[width=\columnwidth]{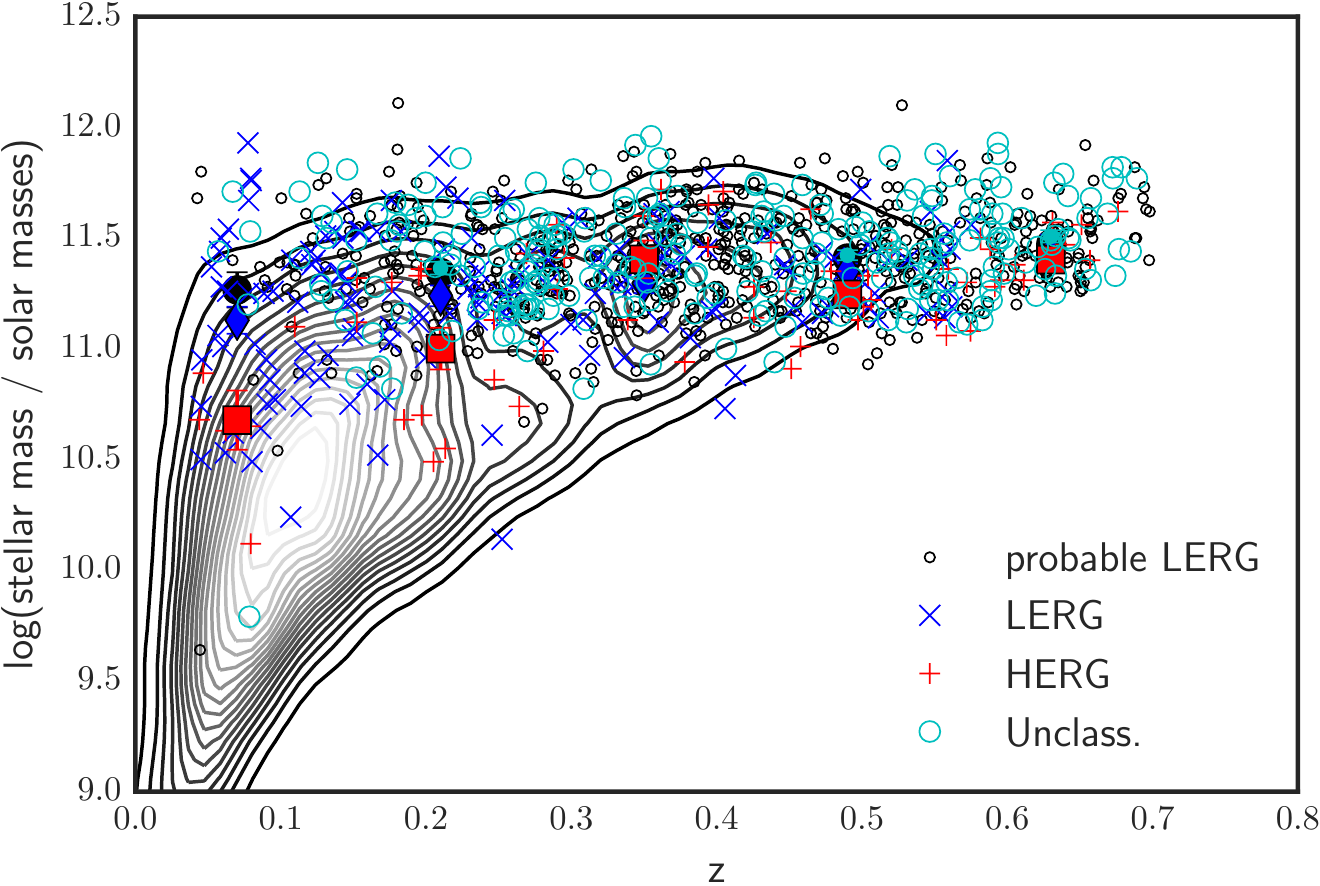}}
\caption{Stellar mass as a function of redshift, with the HERGs, LERGs, probable LERGs and unclassified sources shown separately. The filled shapes show the mean values in each redshift bin for the different samples. The black contours show the distribution of sources in the full SDSS catalogue in the VLA Stripe~82 survey area.}\label{fig:Mstar}
\end{figure}

\begin{figure}
\centerline{\includegraphics[width=\columnwidth]{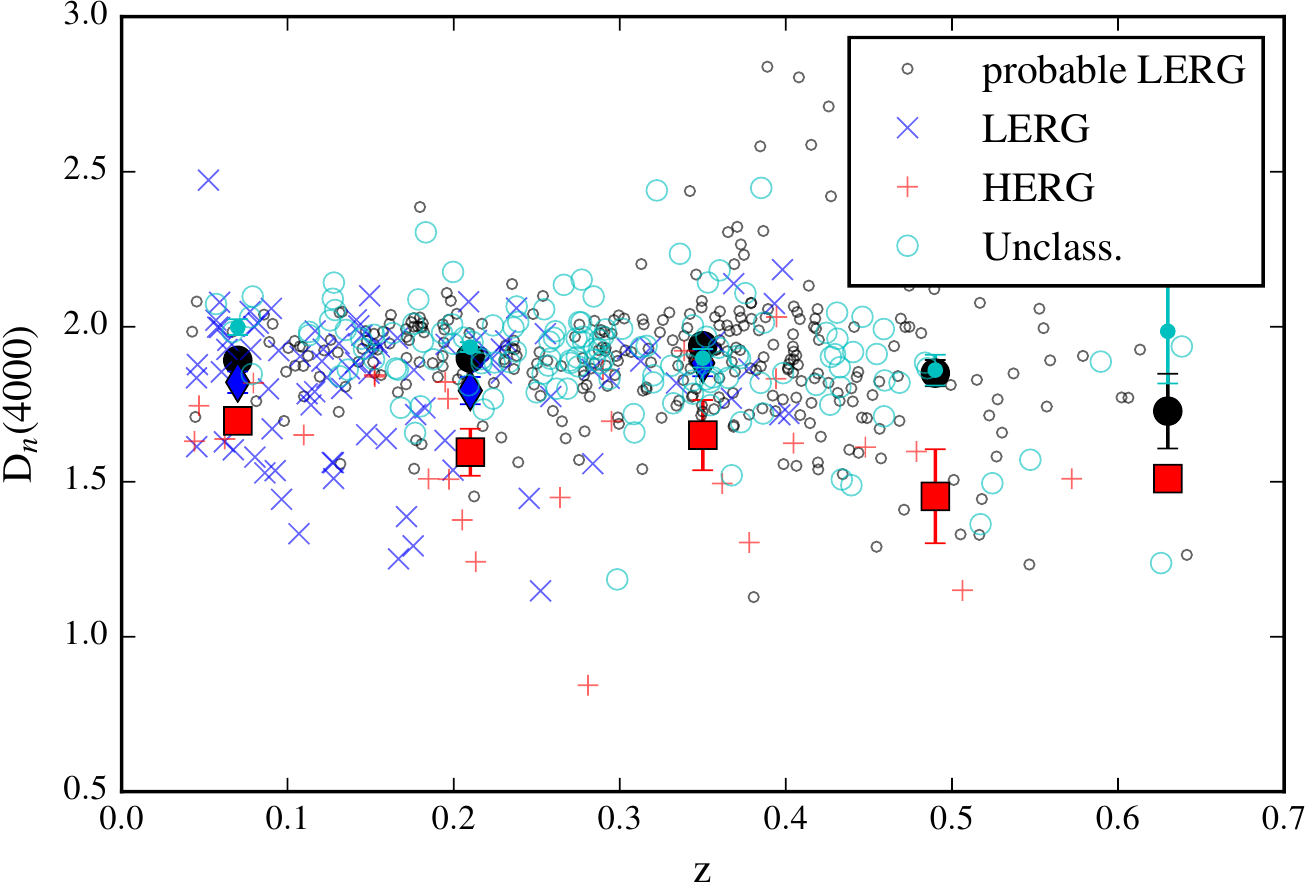}}
\medskip
\centerline{\includegraphics[width=\columnwidth]{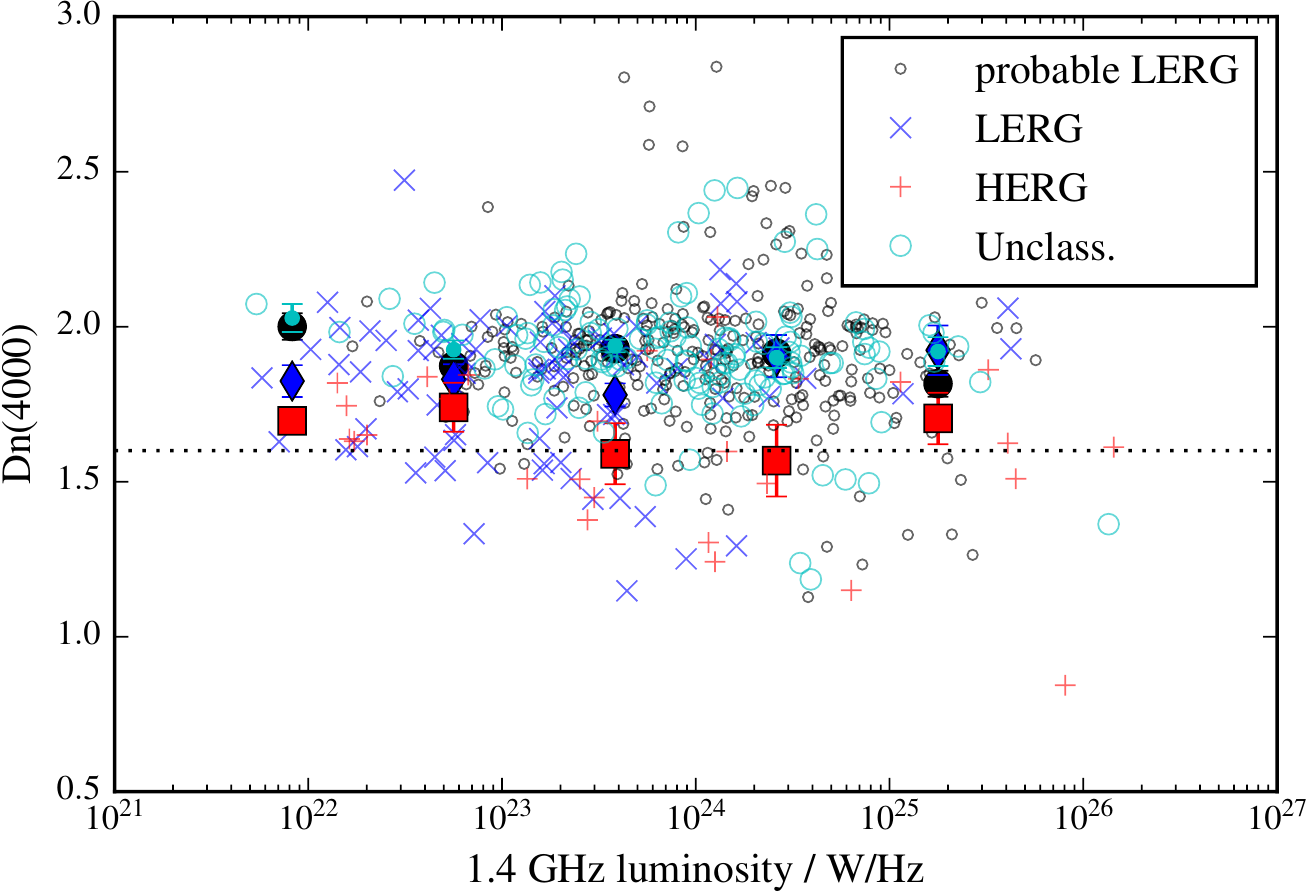}}
\caption{4000~\AA~break strength as a function of redshift (top panel) and 1.4-GHz luminosity (bottom panel), with the HERGs, LERGs, probable LERGs and unclassified sources shown separately. The filled shapes show the mean values in each luminosity bin for the different samples. The dotted line is at D$_{\rm n}$(4000) = 1.6, the value above which galaxies show little evidence of recent star formation \citep{2003MNRAS.341...33K}.}\label{fig:Dn4000}
\end{figure}

Fig.~\ref{fig:Mstar} shows stellar mass as a function of redshift, the contours show the mass and redshift distribution of all SDSS sources in the VLA Stripe~82 survey area for comparison. HERGs have lower stellar masses than than LERGs at redshifts $\lesssim 0.3$, which is consistent with the \citet{2012MNRAS.421.1569B} study in the local Universe which found that HERGs are hosted by lower mass galaxies than LERGs. At higher redshifts ($z \gtrsim 0.3$) the stellar masses of the HERGs and LERGs are very similar; it seems that for the LERGs the stellar mass remains relatively constant across the redshift range, while the HERGs are hosted by more massive galaxies are higher redshifts. However, we are unable to determine whether or not this effect is real due to the limited sensitivity of the SDSS spectra to lower mass objects at $z>0.3$. It is possible that we are only detecting the high-mass tail of the HERG distribution at $z > 0.3$. Deeper spectroscopy, such as that planned by the WEAVE survey of LOFAR fields \citep{2016sf2a.conf..271S}, is required to study the differences in the host masses of HERGs and LERGs at higher redshifts, and study the evolution of their mass functions.

Fig.~\ref{fig:Dn4000} shows the 4000~\AA~break strength as a function of redshift and 1.4~GHz luminosity. 4000~\AA~break strength is a measure of the age of the stellar population present in a galaxy, with younger stellar populations having smaller break strengths. This is because they contain hotter, more massive OB stars which produce multiply-ionised elements, leading to a decrease in opacity. The horizontal dotted line is at D$_{\rm n}$(4000) = 1.6, the value above which galaxies show little evidence of recent star formation \citep{2003MNRAS.341...33K}. There is a clear trend for HERGs to have younger stellar populations than LERGs at all redshifts and luminosities. We can compare this to the work of \citet{2010MNRAS.406.1841H} studying radio galaxies at $z \sim 0.5$, who found two distinct population on the D$_{\rm n}$(4000)-radio luminosity plane; they found that all of the lower-luminosity ($L_{151~{\rm MHz}} < 10^{25.3}~\rm{W \, Hz^{-1}}$) sources in their sample had high 4000~\AA~ break strengths ($>1.6$), suggesting an old stellar population (these sources are also all LERGs, or LEGs in their nomenclature), while the more luminous sources have younger stellar populations. They suggest that this is due to an increased supply of cold gas for star formation in the more luminous sources. This trend is not nearly so evident in our sample - there are a significant number of low-luminosity sources with D$_{\rm n}$(4000) $<1.6$ - and these sources are a mixture of HERGs and LERGs. However, the sources with the lowest D$_{\rm n}$(4000) values, and therefore the youngest stellar populations, do tend to be the more luminous sources. 

While there is a clear trend for the mean D$_{\rm n}$(4000) value to be lower for HERGs than LERGs, there are a significant number of LERGs present in our sample with D$_{\rm n}$(4000)  < 1.6, implying these sources have undergone recent star formation. This suggests that there are galaxies with a large enough reservoir of cold gas to have formed a noticeable fraction of their stellar mass in recent star formation \citep{2003MNRAS.341...33K,2010MNRAS.406.1841H} which are accreting inefficiently, in contrast to the usually-assumed model where LERGs lack a cold gas supply. However, the vast majority of the probable LERGs have D$_{\rm n}$(4000) > 1.6, which implies that there is no evidence for recent star formation in these sources. As mentioned in Section~\ref{section:sample_used}, at $z>0.4$ our sample may be missing some blue objects, due to the colour selection employed by the BOSS survey (where the spectra for the sources in this redshift range originate from). We expect the majority of the bluer radio sources to be SFGs and therefore excluded from the sample discussed in this paper in any case, but there may be some bluer AGN with $z>0.4$ missing from this sample, and therefore not included in Fig.~\ref{fig:Dn4000}. However, we do not expect this selection effect to depend on [OIII] equivalent width and therefore HERG/LERG classification, so it should not affect the results discussed in this section. The lack of any significant difference in D$_{\rm n}$(4000) values between the SDSS and BOSS selected galaxies (with $z<0.4$ and $z>0.4$ respectively) in Fig.~\ref{fig:Dn4000} also suggests that this is not a major effect.

There is a tendency for the properties of the `probable LERGs' to be further from the HERGs than the securely classified LERGs in both stellar mass and D$_{\rm n}$(4000), which suggests that the probable LERGs are perhaps the more extreme tail of the LERG population. This is understandable as these sources are generally not classified using the full criteria because the relevant emission lines are not securely detected, meaning they have the weakest emission lines. The unclassified sources have very similar stellar masses and 4000~\AA~break strengths to the probable LERGs, suggesting that they are drawn from the same population.

\subsection{Mid-infrared properties}\label{section:midIRproperties}

Mid-infrared observations are a useful diagnostic of source type, and can provide valuable information about the properties of both the host galaxy and the AGN. Optical and UV emission from an AGN core is intercepted by the dusty torus, if present, and re-radiated in the mid-infrared. This process is nearly independent of orientation effects \citep{2004ApJ...602..116W,2009ApJ...694..268D}. Mid-infrared observations can therefore be used to identify obscured AGN (e.g.\ \citealt{2004ApJS..154..166L,2005ApJ...631..163S}) and test unification theories of AGN (e.g. \citealt{1992ApJ...391...39H,1994ApJ...428...65H,2005ApJ...629...88S}). Mid-infrared observations can also provide information about recent star-formation via polycyclic aromatic hydrocarbon (PAH) emission.

\begin{figure*}
\centerline{\includegraphics[width=17cm]{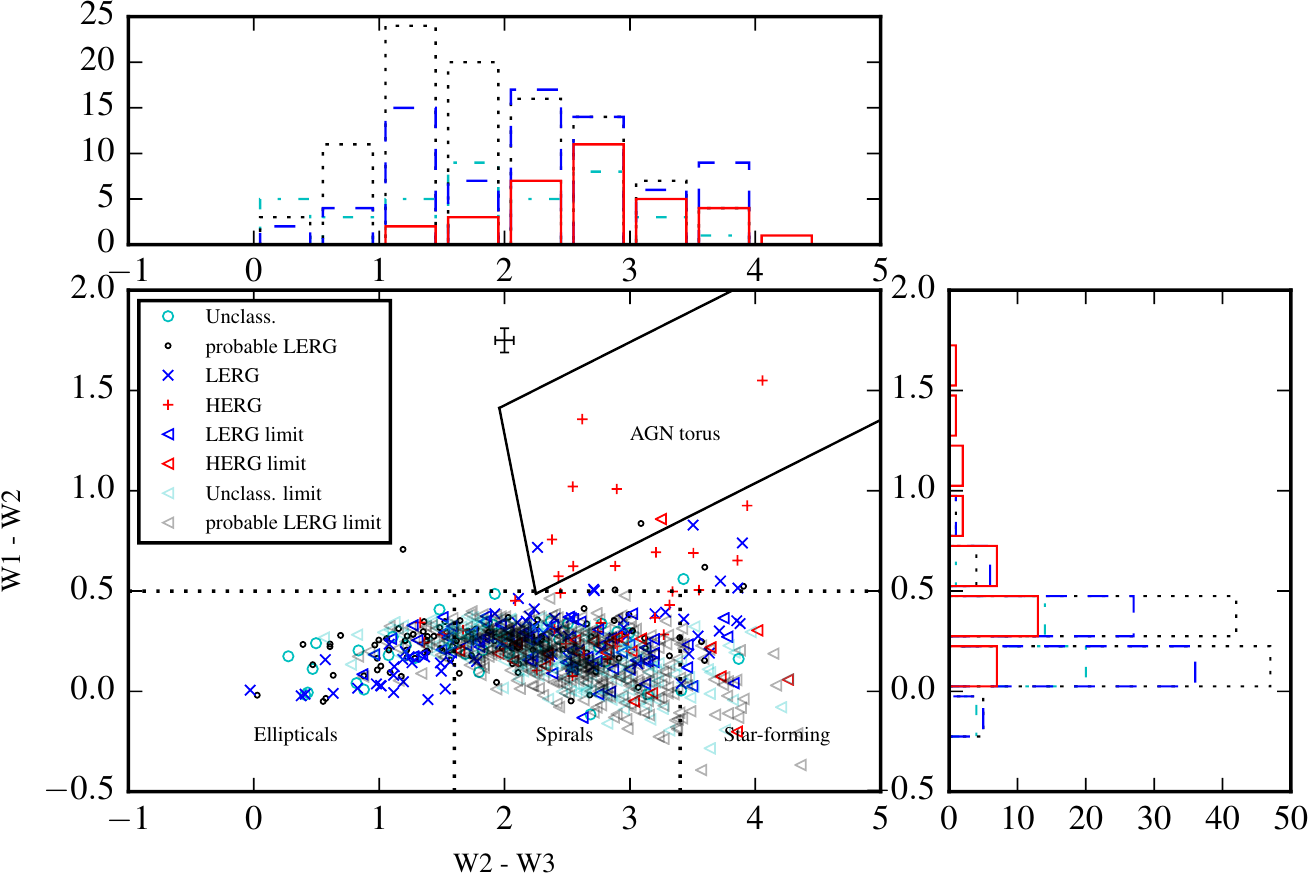}}
\caption{WISE colour-colour diagram. Magnitudes are vega magnitudes.  Sources with upper limits on their W3 magnitudes are plotted as triangles and could move to the left.  HERGs, LERGs, probable LERGs and unclassified sources are shown separately. The black solid line indicates the three-band AGN selection from \citet{2012MNRAS.426.3271M} in which objects with a dusty AGN torus are expected to lie. The three dotted boxes in the bottom half of the diagram are the regions in which ellipticals, spirals and star-forming galaxies (from left to right respectively) are expected to be found from \citet{2016MNRAS.462.2631M}. An example point is plotted with error bars showing the size of the median uncertainties in the points plotted.
}\label{fig:WISEcolours}
\end{figure*}

We are able to study the infrared properties of the sources in this sample using data from the \emph{WISE} survey. The \emph{WISE} data products used here and the procedure used to match the \emph{WISE} catalogue to our sample are described in Section~\ref{section:WISE_data}. The \emph{WISE} colour-colour diagram, (W1 - W2) against (W2 - W3), is a particularly useful diagnostic diagram, as the (W1 - W2) colour traces the fractional contribution of non-stellar emission from an AGN and the (W2 - W3) colour is related to the specific star-formation rate and the contribution from warm dust \citep{2010AJ....140.1868W,2012ApJ...748...80D,2017MNRAS.464.1306C}. Different source types therefore typically lie in different locations in this diagram. \citet{2016MNRAS.462.2631M} developed a series of colour cuts of this diagram, based on the work of \citet{2012AJ....143....7L}, which separate elliptical, spiral and starburst galaxies from AGN. (These are shown by the dotted lines in Fig.~\ref{fig:WISEcolours}). Note, however, that they find significant overlap between the different populations in these diagrams. This diagram has also been widely used to separate AGN from star-forming galaxies (e.g.\ \citealt{2012MNRAS.426.3271M}), although it has been shown that this method misses many AGN, particularly LERGs (e.g.\ \citealt{2014MNRAS.438.1149G}), as only AGN with power-law emission from a dusty torus will be selected. For objects with an AGN torus, the torus emission can be considered a proxy for intrinsic AGN luminosity (e.g.\ \citealt{2011MNRAS.411.1909F}), so can be used to investigate the power of these AGN.

\begin{figure}
\centerline{\includegraphics[width=\columnwidth]{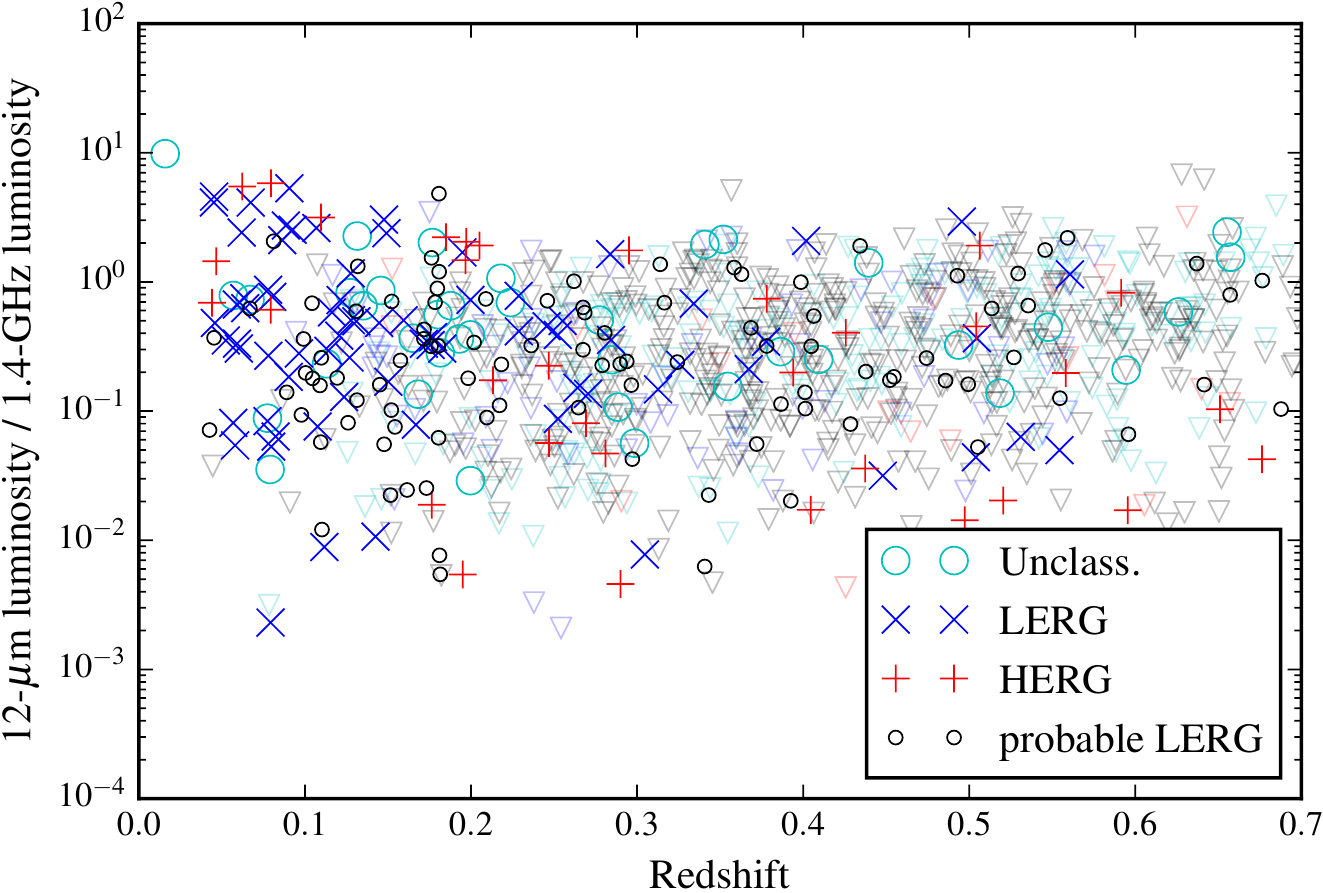}}
\caption{The ratio between rest frame 12-$\muup$m luminosity and 1.4-GHz luminosity as a function of redshift, with HERGs, LERGs, probable LERGs and unclassified sources shown separately. Sources with an upper limit in the W3 band are shown as triangles.}\label{fig:L12L14z}
\end{figure}

\begin{figure}
\centerline{\includegraphics[width=\columnwidth]{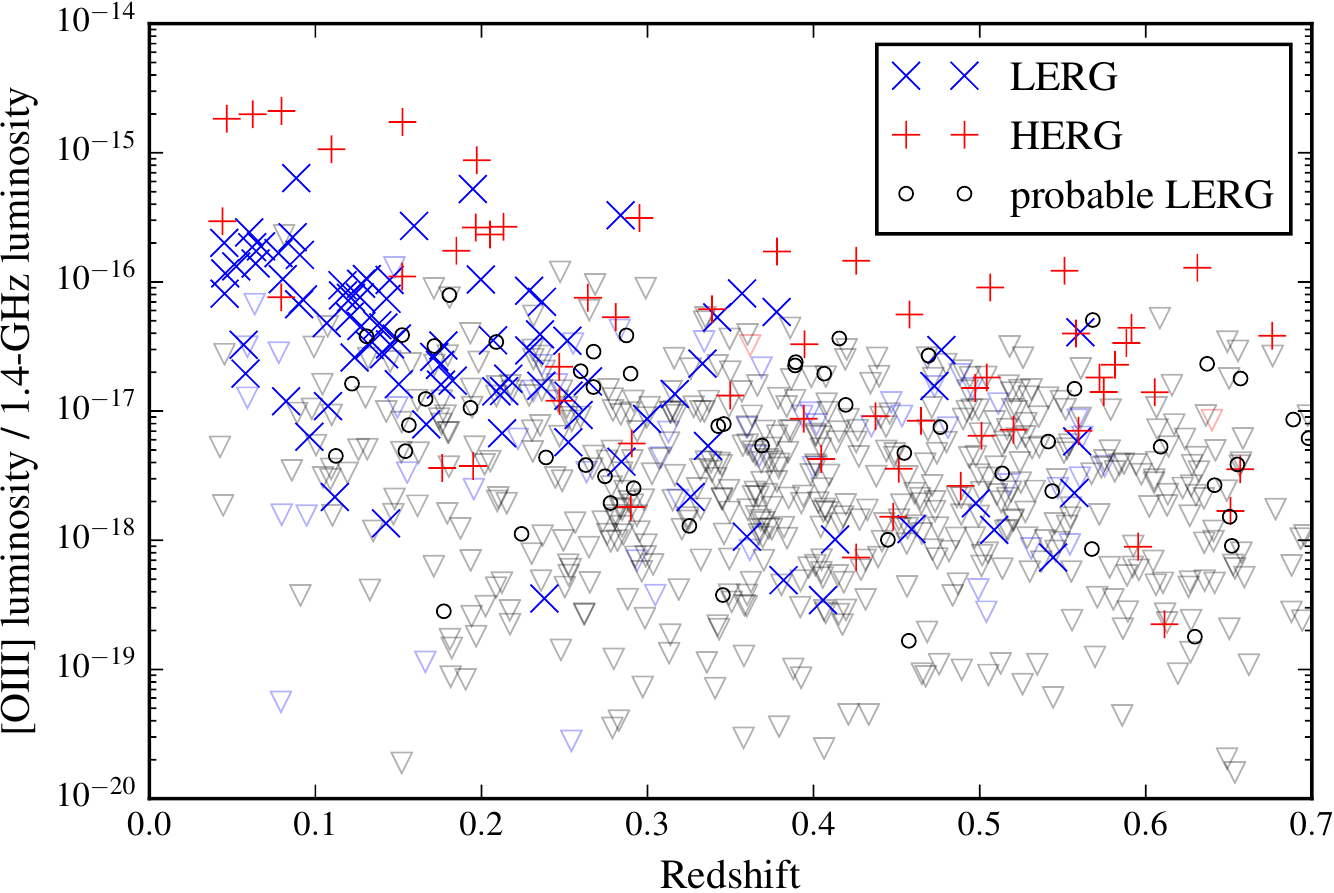}}
\caption{The ratio between [OIII] line luminosity and 1.4-GHz luminosity as a function of redshift, with HERGs, LERGs and probable LERGs shown separately. Sources with an upper limit on [OIII] luminosity are shown as triangles. The unclassified sources do not have an [OIII] line measurement so are unable to be included in this Figure.}\label{fig:LOIIIL14z}
\end{figure}

Fig.~\ref{fig:WISEcolours} shows the \emph{WISE} colour-colour diagram using W1, W2 and W3 (3.4, 4.6 and 12-$\muup$m) magnitudes. The region in which sources with a dusty AGN torus are expected to be found is marked by solid black lines. This region mostly only contains HERGs, as expected, with the exception of one LERG and one `probable LERG' towards the edge of this region, where the associated uncertainties could move them out of the region. There are, however, many HERGs found outside this region, indicating that they do not have significant emission from an AGN torus, and that a large proportion of HERGs would be missed if this region is used to select HERGs. 

LERGs are mostly only found in the elliptical and spiral regions of the diagram, suggesting they are dominated by old stars with an increasing contribution from warm dust as you move to redder W2 - W3 colours. HERGs are found scattered across the whole diagram, which is a similar result to that found by \citet{2017MNRAS.464.1306C}, who postulate that this indicates that HERGs are a heterogeneous class of objects, where the physical process causing the mid-infrared emission is not the same for all objects. Such a scatter could, however, also be caused by orientation effects, which are expected to affect HERGs as they are thought to have a dusty torus. There are, however, very few HERGs in the `ellipticals' region in the bottom left of the diagram, corresponding to the region where sources with the lowest specific star-formation rates and smaller contribution from dust may be found. This is consistent with the result from Section~\ref{section:galproperties} that HERGs tend to have higher star-formation rates than LERGs. 
The unclassified sources are mostly found in the elliptical and spiral regions of the diagram in a similar way to the LERGs and probable LERGs.

Fig.~\ref{fig:L12L14z} shows the ratio between infrared and radio luminosities as function of redshift. The 12-$\muup$m luminosity is a proxy for the radiative power of AGN which have a dusty torus, and the 1.4-GHz luminosity relates to the jet power. We might therefore expect the HERGs to have higher 12-$\muup$m luminosities  compared to their radio luminosities than the HERGs (see e.g.\ the \citealt{2014MNRAS.438.1149G} study at higher radio luminosities), however there is no noticeable difference between the HERGs and the LERGs in this diagram. This may be because a significant number of sources are undetected in the W3 band, particularly at $z > 0.3$, so only upper limits are available. It is, however, notable that for those sources with detections both the HERGs and LERGs are found across the full range of $L_{22~\muup m} / L_{1.4~\rm GHz}$ ratios. Several studies (e.g.\ \citealt{2007ApJ...660..117C,2009MNRAS.396.1929H,2014MNRAS.438.1149G}) have found that there is a correlation between $L_{12~\muup\rm{m}}$ and $L_{1.4~\rm GHz}$ for HERGs but not for LERGs, which does not seem to be the case here as the two populations are similarly distributed.  

The mid-infrared luminosity and the [OIII] line luminosity both trace the radiative power of the AGN, as the AGN emission excites the gas clouds in the narrow line region and heats the circumnuclear dust. We therefore show the ratio of [OIII] line luminosity to radio luminosity as a function of redshift in Fig.~\ref{fig:LOIIIL14z} for comparison with Fig.~\ref{fig:L12L14z}. There is a tendency for the HERGs to have higher [OIII] luminosities compared to their radio luminosities, which is consistent with the scenario where they radiate efficiently across the electromagnetic spectrum, while LERGs emit the bulk of their power in kinetic form as radio jets (e.g.\ \citealt{2007MNRAS.381..589M,2014ARA&A..52..589H}). Again there is no evidence that the radiative and mechanical powers are more correlated for HERGs than for LERGs.

The scatter in the ratios of both $L_{22~\muup m}$ and $L_{\textrm{[OIII]}}$ to $L_{1.4~\rm GHz}$ of $\gtrsim 3$~dex indicates that there is not a one-to-one relationship between radiative and jet powers. This is broadly in agreement with the work by \citet{2016MNRAS.462.2631M} at higher radio luminosities who found a scatter of $\sim 4$~dex between the radiative and kinetic output of the sources in their sample, but is in disagreement of earlier work by \citet{1991Natur.349..138R} and \citet{1999MNRAS.309.1017W} who find a tighter relationship at higher radio luminosities.

\section{Accretion rates}\label{section:rates}

\subsection{Accretion rates of HERGs and LERGs}\label{section:ratesHERGs}

\begin{figure}
\centerline{\includegraphics[width=\columnwidth]{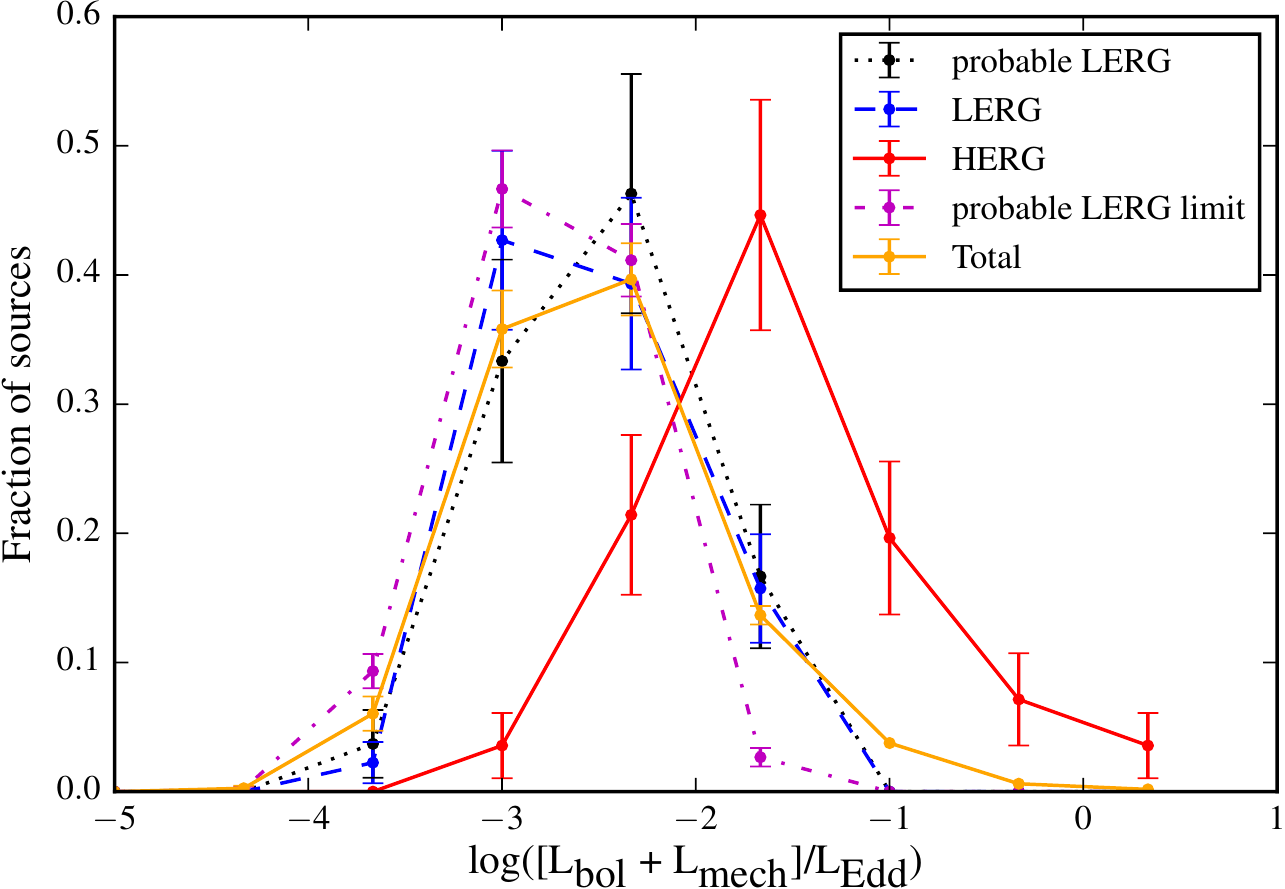}}
\medskip
\centerline{\includegraphics[width=\columnwidth]{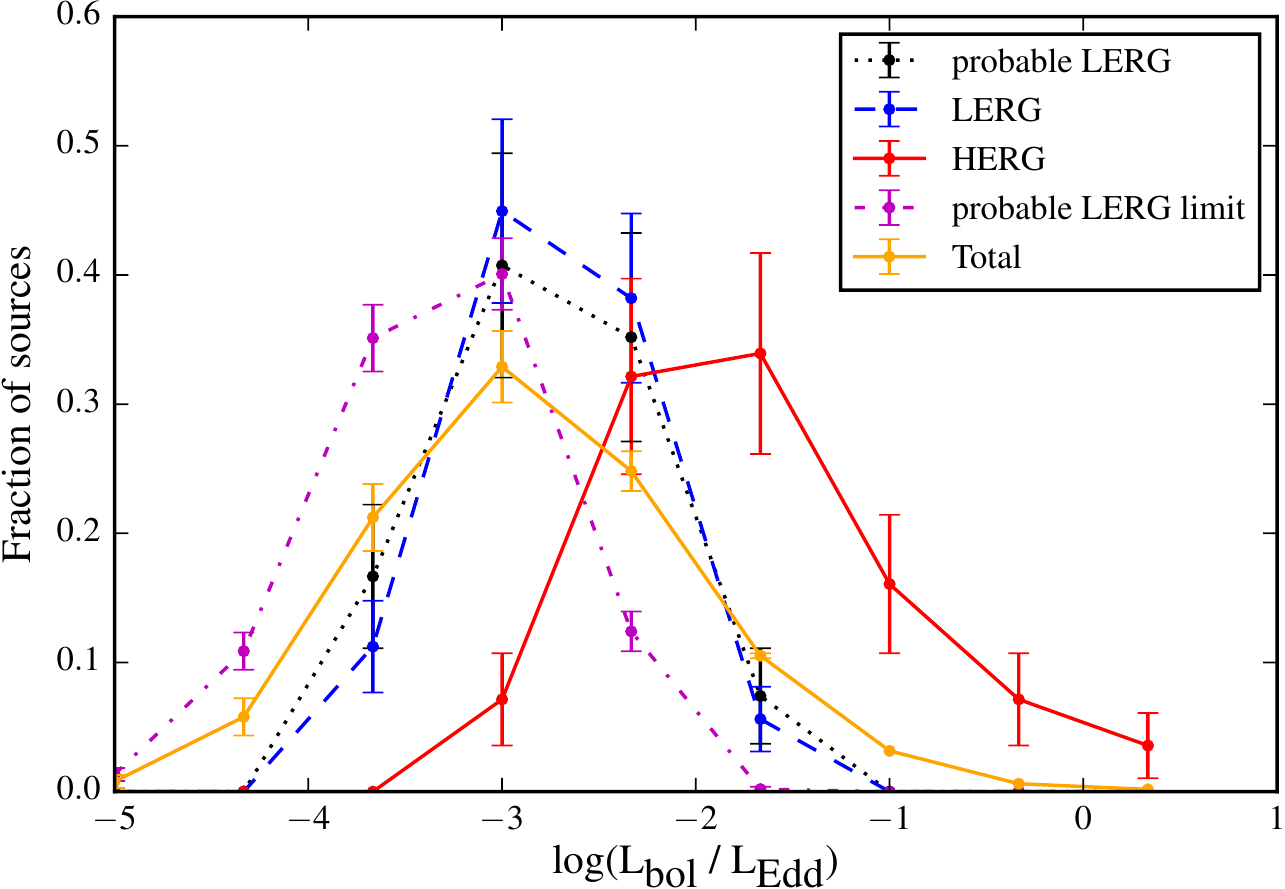}}
\caption{Distributions of Eddington-scaled accretion rates for the different source classifications. Top panel shows the combined radiative and mechanical luminosity (see text for details) while the bottom panel shows the radiative accretion rate only. }\label{fig:accretion}
\end{figure}

\begin{figure}
\centerline{\includegraphics[width=\columnwidth]{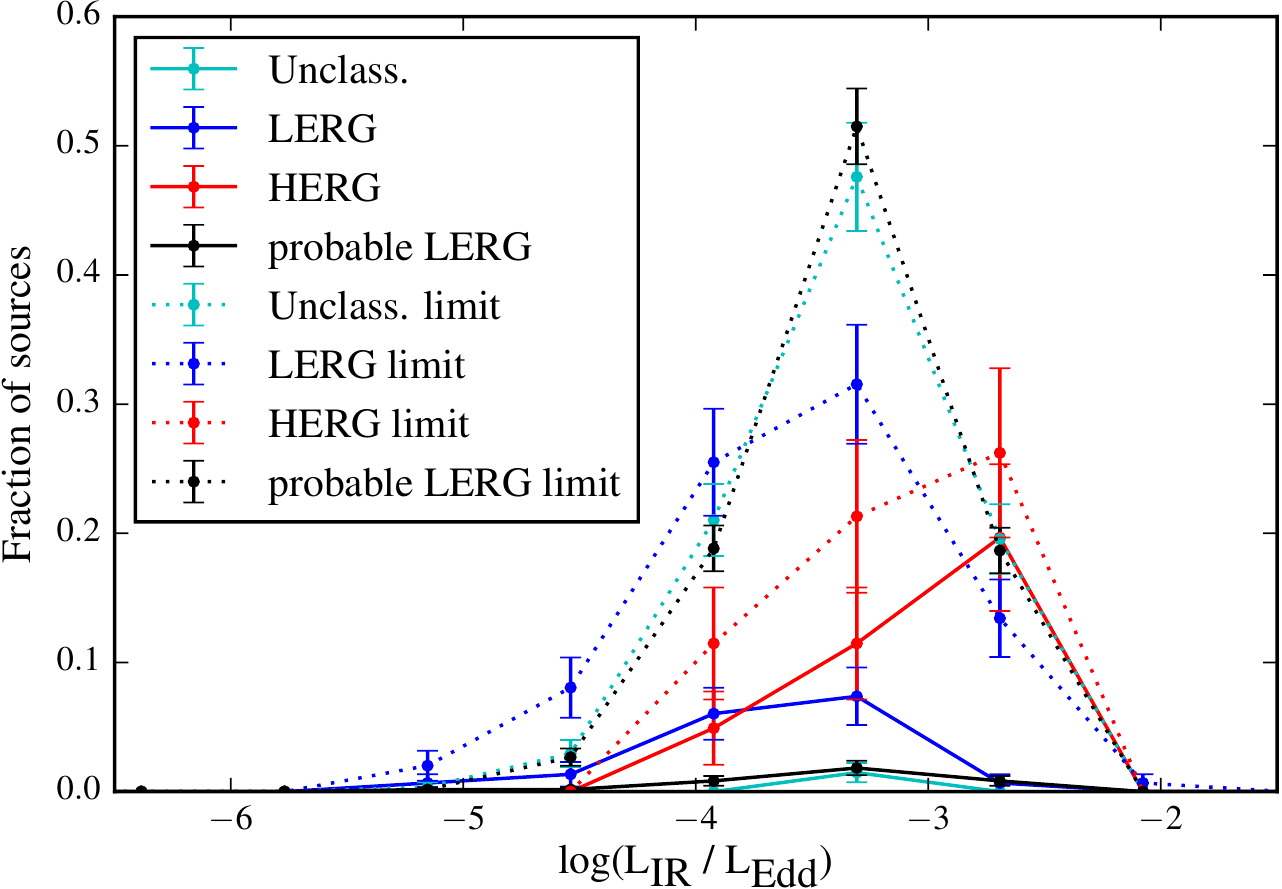}}
\caption{Distribution of 12-$\muup$m luminosity / Eddington luminosity (a proxy for the Eddington-scaled radiative accretion rate) for the difference source classifications. The distributions for sources with a 12~$\muup$m detection are shown as solid lines, and those with an upper limit in the 12-$\muup$m band are shown as dotted lines.}\label{fig:accretionLir}
\end{figure}

The scenario building up in the literature (see review by \citealt{2014ARA&A..52..589H} and references therein) is that the different properties of HERGs and LERGs are related to the fact that they accrete matter at different rates.
This idea can be explored further by calculating the Eddington-scaled accretion rates for the sources in this sample and investigating how these relate to the host galaxy properties discussed in Section~\ref{section:properties}.

The radiative luminosity ($L_{\rm bol}$) of each source was estimated from the [OIII] 5007 emission line luminosity ($L_{\rm [OIII]}$) using the following expression derived by \citet{2004ApJ...613..109H}: $L_{\rm bol} = 3500 L_{\rm{[OIII]}}$. The uncertainty in this relation is 0.4~dex. For sources with an AoN < 1.5 their [OIII] line measurement is considered an upper limit on their radiative luminosity. 

The mechanical luminosity of the radio jet ($L_{\rm mech}$) was estimated from the 1.4~GHz luminosity using the relationship from \citet{2010ApJ...720.1066C}, $L_{\rm mech} = 7.3 \times 10^{36} (L_{1.4~\rm{GHz} }/ 10 ^{24} \rm{W~Hz^{-1}})^{0.7}$~W, which has a scatter of 0.7~dex. The scatter in this relationship may be due to differing particle compositions of the radio jets; sources with a larger fraction of non-radiating particles will have a higher jet power for a given radio luminosity. A study of the particle content of radio galaxies by \citet{2018MNRAS.tmp..274C} found no systematic difference between HERGs and LERGs (although they did find a difference for different radio galaxy morphologies), suggesting that any difference seen in the radio luminosities of HERGs and LERGs translates into a difference in jet powers, rather than being due to a different scaling relationship for the two populations. The \citet{2010ApJ...720.1066C} relation used in this work is consistent with the \citet{1999MNRAS.309.1017W} scaling relation if one assumes that the ratio between non-radiating particles and relativistic electrons is several tens of thousands (consistent with results in the literature, e.g.\ \citealt{2005MNRAS.364.1343D,2006MNRAS.372.1741D,2008ApJ...686..859B}).
The black hole masses were estimated from stellar masses using the local black hole mass - bulge mass relation: $M_\textrm{BH} \sim 0.0014M_{\ast}$ \citep{2004ApJ...604L..89H}.
The Eddington limit for each source was then calculated as follows: $L_{\textrm{Edd}} = 1.3 \times 10^{31} M_{\textrm{BH}} / M_\odot \textrm{W}$. 

The Eddington-scaled accretion rate for each source was calculated as $\lambda = (L_{\rm bol} + L_{\rm mech}) / L_{\rm Edd}$; the top panel of Fig.~\ref{fig:accretion} shows the distribution of these accretion rates for the different groups of sources. The HERGs generally have higher accretion rates than the LERGs, with the distribution peaking just below 0.1 compared to 0.01 for the LERGs. However, there is a considerable overlap between the two populations, with HERGs found across almost the full range of accretion rates. 
The probable LERGs show similar Eddington-scaled accretion rates to the securely classified LERGs, although as many of these values are upper limits their accretion rates may in fact be lower. As we do not have [OIII] line measurements for the unclassified sources we are unable to calculate their radiative accretion so they are not included in this section.

The Eddington-scaled accretion rates can be calculated considering the radiative luminosities only (and therefore ignoring the mechanical luminosity in the jets) as follows: $L_{\rm bol} / L_{\rm Edd}$. These are shown in the bottom panel of Fig.~\ref{fig:accretion}. Again the LERGs show lower accretion rates than the HERGs but with some overlap between the populations. Taking into account the upper limits, the probable LERGs show lower radiative-only accretion rates than both the HERGs and the securely classified LERGs. This is consistent with the fact that the host galaxies of these probable LERGs have larger masses and older stellar populations than the classified LERGs (see Section~\ref{section:galproperties}). This could imply that there is a continuous range of accretion rates, resulting in a continuous range of host galaxy properties. This possibility is explored further in Section~\ref{section:accretion_properties}. 

As described in Section~\ref{section:midIRproperties}, the mid-infrared luminosity of a source can also provide a measure of the accretion rate. We therefore use $L_{12~\muup\textrm{m}} / L_\textrm{Edd}$ as a proxy for the Eddington-scaled accretion rate. 
The distribution of the Eddington-scaled luminosities calculated in this way for HERGs and LERGs are shown in Fig.~\ref{fig:accretionLir}. Despite large number of upper limits on the infrared luminosities (due to non-detections in the \emph{WISE} W3 band) these results suggest that the HERGs have higher radiative accretion rates than the LERGs but that there is a considerable overlap between the two distributions. This supports the results using [OIII]. Due to the relative depths of the two datasets used there are significantly more upper limits when using \emph{WISE} compared to [OIII], which is responsible the HERGs and LERGs appearing less separated in Fig.~\ref{fig:accretionLir} than in Fig.~\ref{fig:accretion}. 12-$\muup$m detections or upper limits are available for the unclassified sources so they are able to be included in this Figure. Although most of the sources have upper limits, they seem to display a very similar distribution to the probable LERGs.

\begin{figure}
\centerline{\includegraphics[width=\columnwidth]{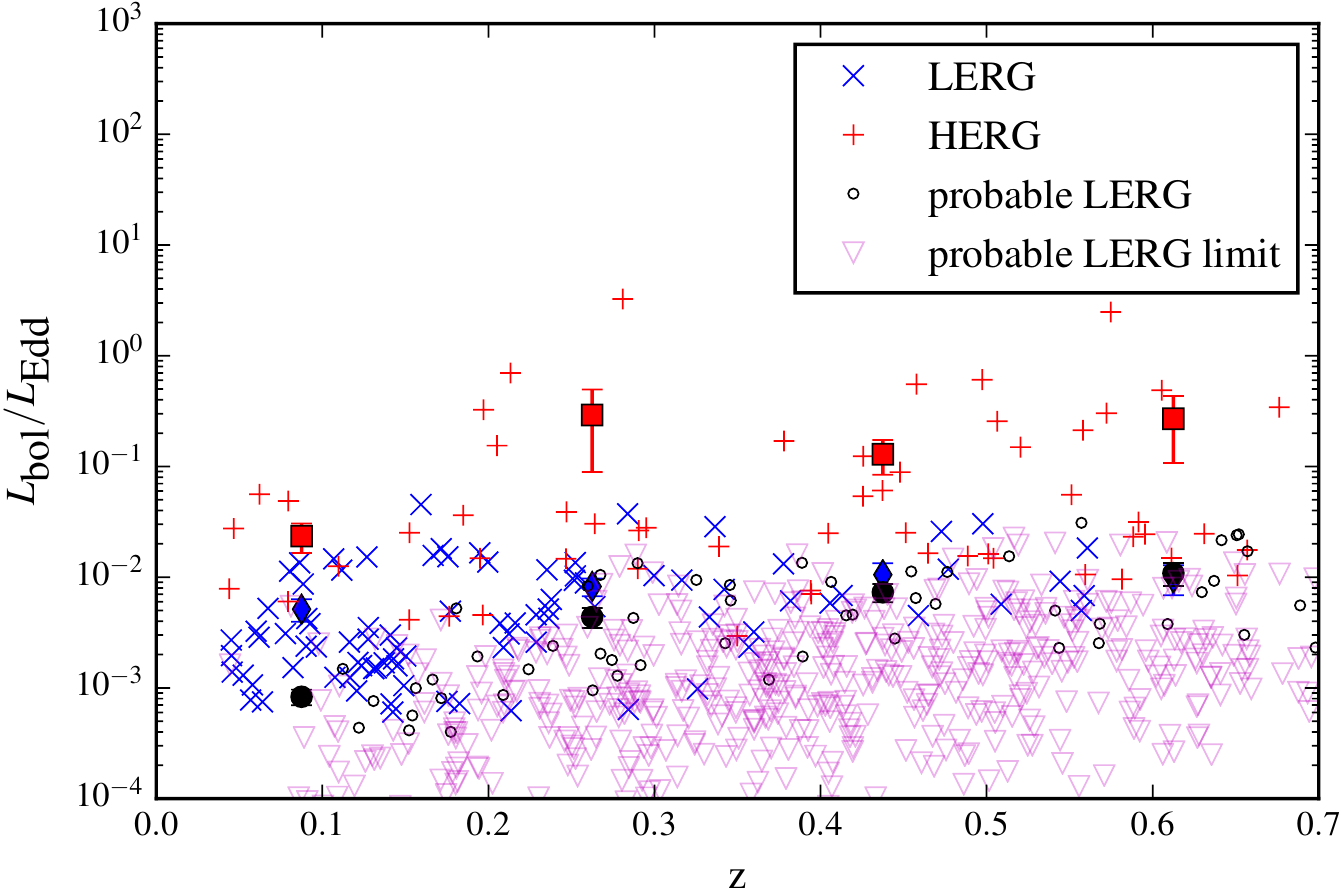}}
\medskip
\centerline{\includegraphics[width=\columnwidth]{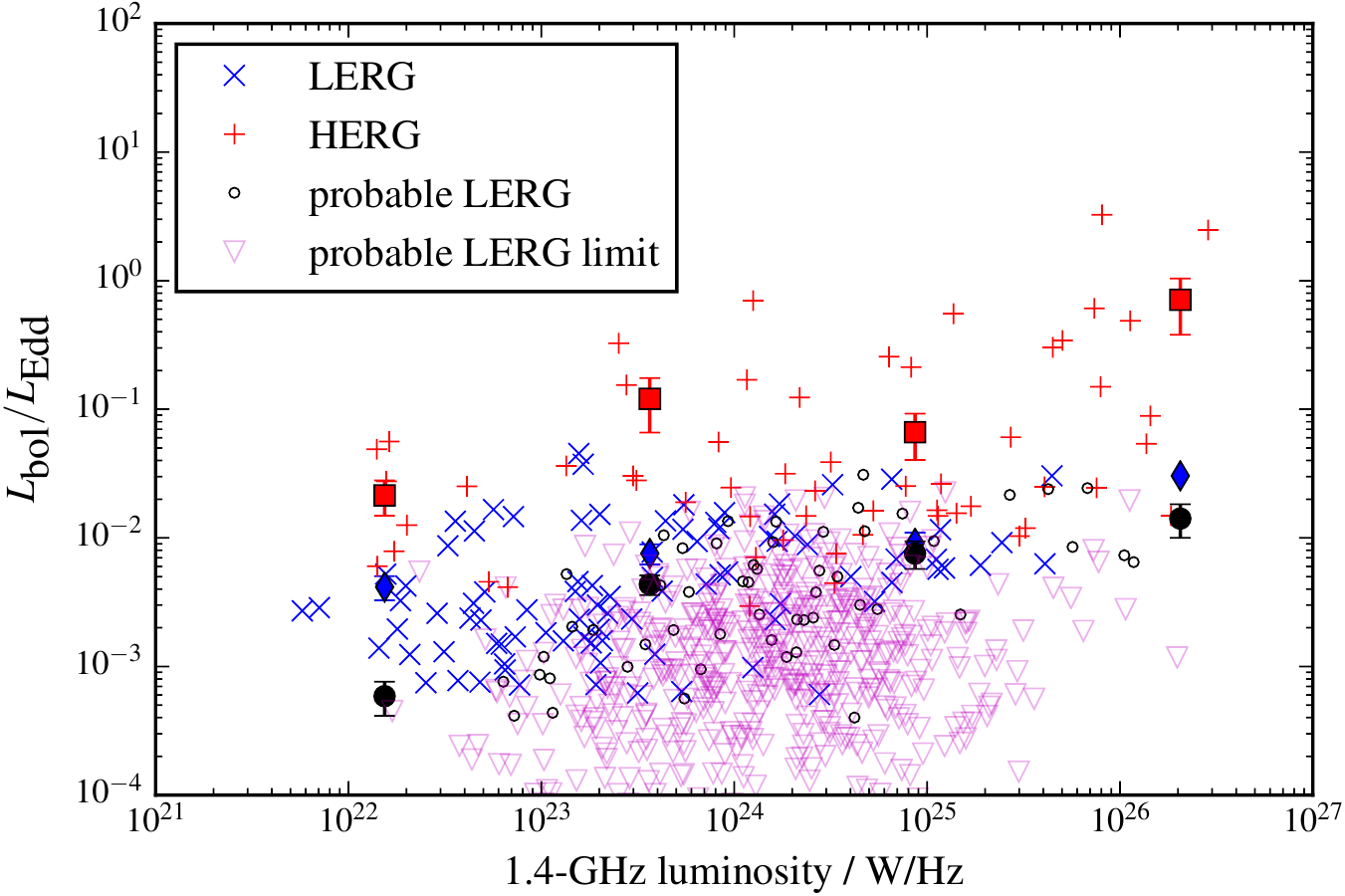}}
\caption{Eddington-scaled accretion rates as a function of redshift (top panel) and 1.4-GHz luminosity (bottom panel) for the different source types. The filled shapes show the mean accretion rate and its uncertainty in each luminosity/redshift bin. Means are not plotted for the limits. The uncertainties in the scaling relations used to estimate $L_\textrm{bol}$ and $L_\textrm{mech}$ are 0.4 and 0.7 dex respectively.}\label{fig:accretionzL14}
\end{figure}

Fig.~\ref{fig:accretionzL14} shows the Eddington-scaled radiative accretion rates as a function of redshift and radio luminosity. There is very little change with redshift, although the overlap in accretion rates between the populations is most noticeable at $z \lesssim 0.2$, due to the lack of sources with the highest accretion rates in this redshift range.
There is some indication that the radiative accretion rate for HERGs increases as 1.4-GHz luminosity increases, while the radiative accretion rate for LERGs seems to be independent of radio luminosity. 
This suggests that LERGs are all accreting at a low rate and are launching radio jets with a range of different powers that is not strongly correlated with accretion rate. For HERGs, however, the power of the radio jet is related to the accretion rate, with the sources accreting at the highest rate having the largest radio luminosities. This supports previous studies using more luminous samples e.g.\ \citet{1999MNRAS.309.1017W}.

The transition between the HERG and LERG populations seems to occur at an accretion rate of $\sim 0.01$, and remains approximately constant with redshift and luminosity. This is broadly similar to the switch between populations reported in other work \citep{2012MNRAS.421.1569B,2014MNRAS.440..269M,2015MNRAS.447.1184F}, and is in agreement with theoretical predictions of where the accretion rate becomes radiatively inefficient (e.g.\ \citealt{1982Natur.295...17R,1995ApJ...452..710N,1997ApJ...489..865E}). However, there is more overlap between the two distributions in this work than reported in other studies. 
\citet{2012MNRAS.421.1569B} found that the distribution of accretion rates for HERGs and LERGs were almost completely separate, with the very small overlap between the two distributions attributed to possible uncertainties in the scaling relation used to estimate $L_{\rm mech}$ (which is the same as used in this work). Note that their analysis is limited to $z < 0.1$, although we see significant overlap in this redshift range in our sample, as discussed. \citet{2015MNRAS.447.1184F} also report a bimodality in their study at $z\sim1$, with a division at approx $\lambda \sim 0.04$, although they have a relatively small sample (27 sources), and there does seem to be some overlap in their sample. 
\citet{2014MNRAS.440..269M} also find a clear bimodality in the accretion rate distribution of their sample of 2~Jy and 3CRR with redshifts in the range $0.02 < z < 0.7$. They interpret this as evidence for a `switch' between accretion modes at a certain value of $L/L_{\rm Edd}$. They do, however, report some overlap in accretion rates between the two populations, which they attribute to the black hole masses of LERGs in rich clusters potentially being underestimated. Note that the sources in the \citeauthor{2014MNRAS.440..269M} and \citet{2015MNRAS.447.1184F} samples generally have higher radio luminosities than the sources in our sample, with luminosity ranges of $10^{25} \lesssim L_{1.4~\textrm{GHz}} / \textrm{W Hz}^{-1} \lesssim 10^{29}$ and of $10^{26} \lesssim L_{1.4~\textrm{GHz}} / \textrm{W Hz}^{-1} \lesssim 10^{29}$ respectively (luminosities are scaled to 1.4 GHz by assuming a spectral index of 0.7), so it is possible that while a bi-modality exists at higher luminosities the situation is more complex at lower radio luminosities.

%------------------------------------------------------------------------------%
\subsection{Galaxy properties as a function of accretion rate}\label{section:accretion_properties}

\begin{figure}
\centerline{\includegraphics[width=\columnwidth]{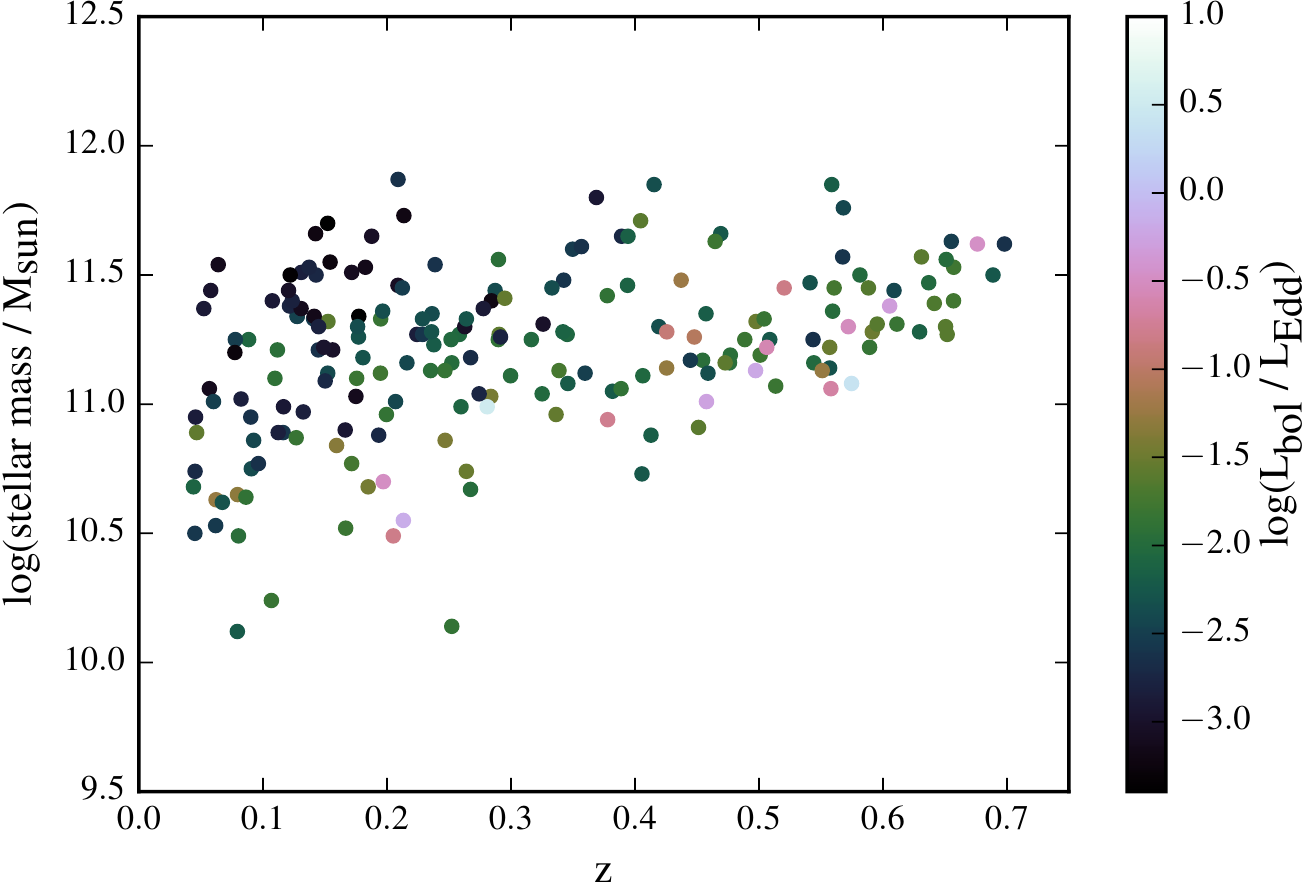}}
\medskip
\centerline{\includegraphics[width=\columnwidth]{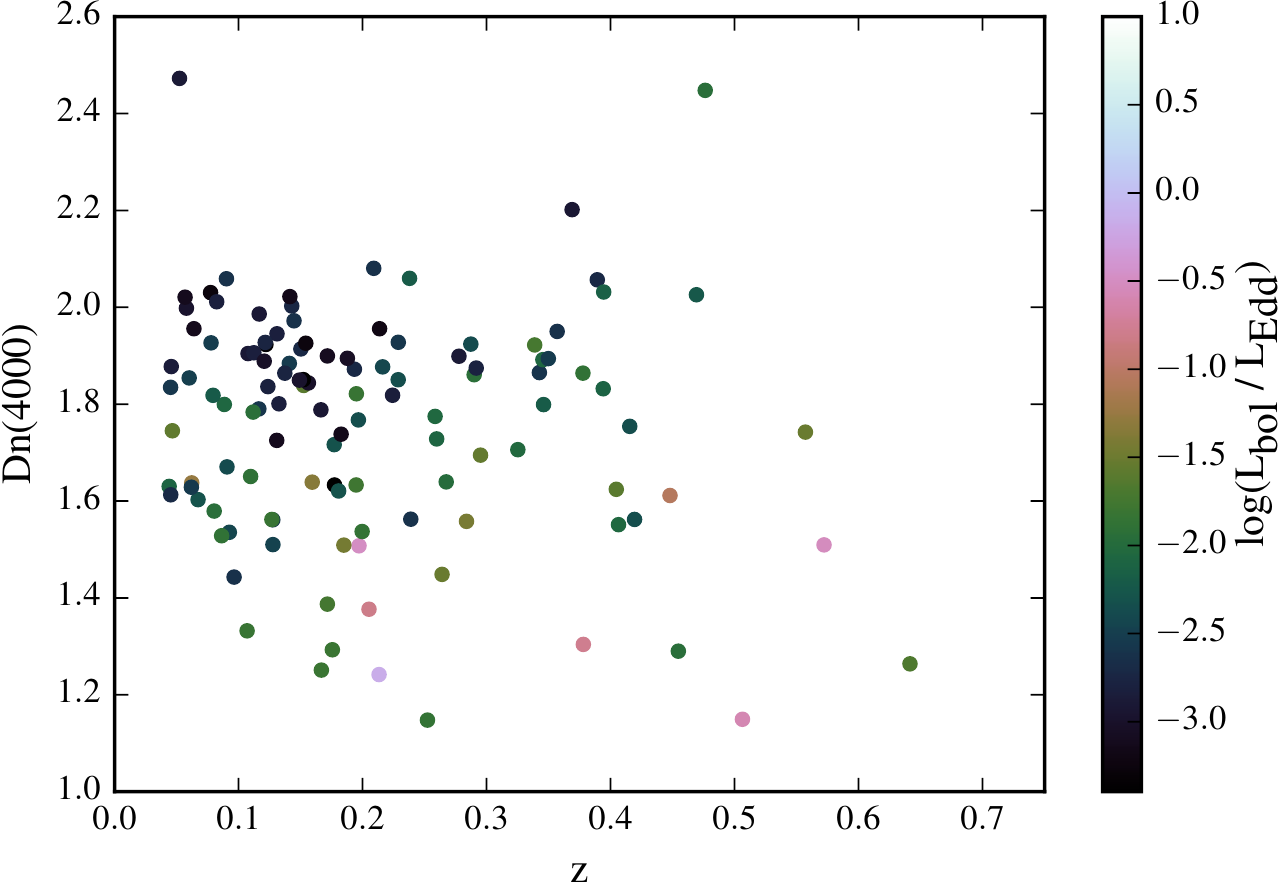}}
\caption{Host galaxy stellar mass and 4000~\AA~break strength as a function of redshift coloured according to Eddington-scaled radiative accretion rate.}\label{fig:propertiesz_colour}
\end{figure}

\begin{figure}
\centerline{\includegraphics[width=\columnwidth]{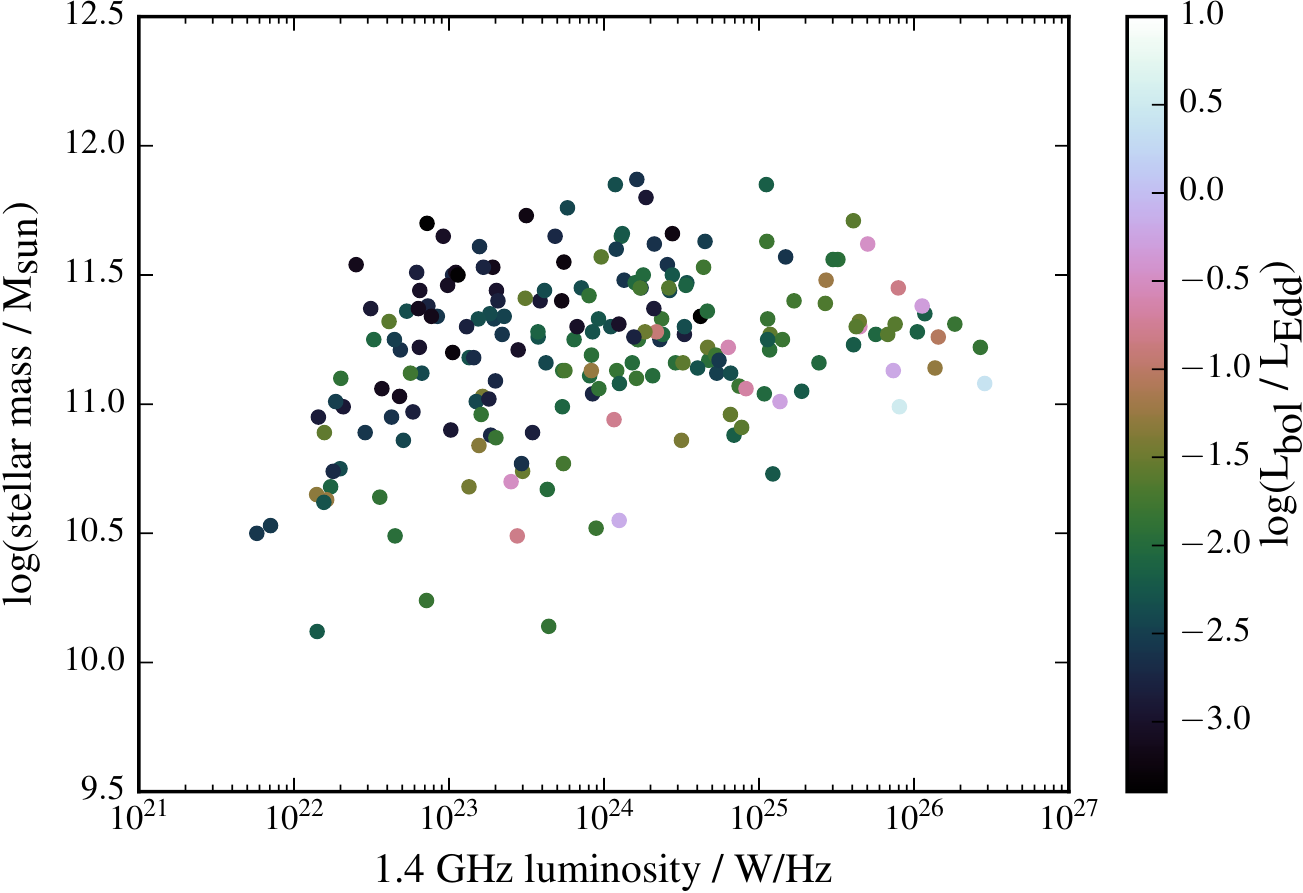}}
\medskip
\centerline{\includegraphics[width=\columnwidth]{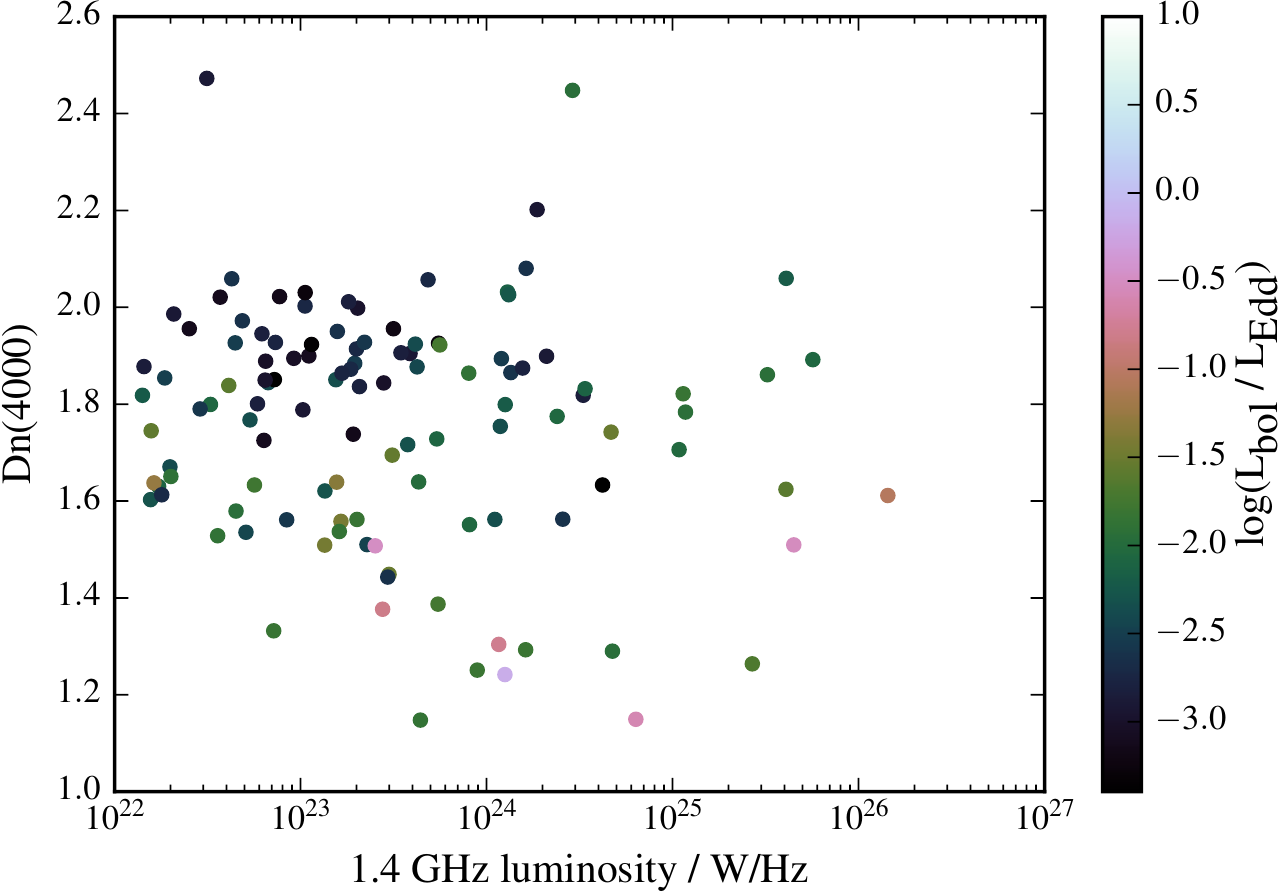}}
\caption{Host galaxy stellar mass and 4000~\AA~break strength as a function of 1.4~GHz radio luminosity coloured according to Eddington-scaled radiative accretion rate.}\label{fig:propertiesL14_colour}
\end{figure}

\begin{figure}
\centerline{\includegraphics[width=\columnwidth]{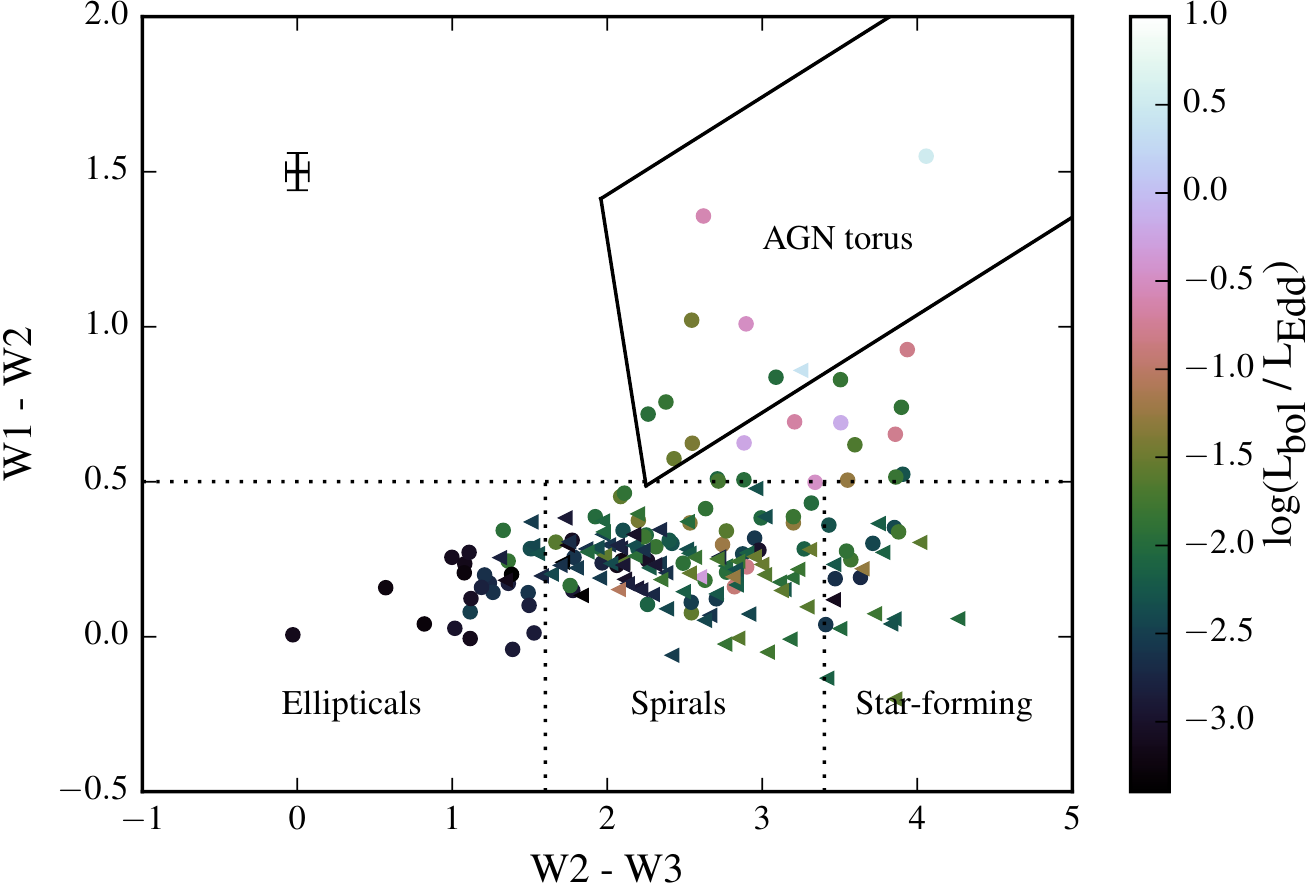}}
\caption{WISE colour-colour diagram. The black solid line indicates the three-band AGN selection from \citet{2012MNRAS.426.3271M} in which objects with a dusty torus are expected to lie, and the dotted lines show the \citet{2016MNRAS.462.2631M} regions in which different source types are expected to be found, as labelled. Sources with upper limits on their W3 magnitudes are plotted as triangles and could move to the left. An example point is plotted with error bars showing the size of the median uncertainties in the points plotted. Sources are coloured according to their Eddington-scaled radiative accretion rate.
}\label{fig:WISEcolours_colour}
\end{figure}

In order to explore the possibility that this sample of radio sources is better described by a continuous distribution of accretion rates, rather than as two discrete populations, we reproduce several of the diagnostic plots discussed earlier in this paper with the sources coloured according to their accretion rate rather than split into HERGs and LERGs. 

The plots of host galaxy properties as a function of redshift, shown in Fig.~\ref{fig:propertiesz_colour}, show that D$_\textrm{n}$(4000) and stellar mass vary smoothly with accretion rate, particularly at low redshift ($z \lesssim 0.3$), where we are sensitive to the lowest radio luminosities. The most slowly accreting sources are found in the most massive galaxies with the oldest stellar populations. At $z > 0.3$ there are fewer sources with the lowest accretion rates, likely due to selection effects, but the general trend remains the same for stellar age, with perhaps a greater scatter in accretion rates than at lower redshift. For stellar mass, however, there is a much smaller spread at higher redshifts (due to selection effects), and these source show a range of accretion rates. 
Fig.~\ref{fig:propertiesL14_colour} shows the galaxy properties as function of 1.4-GHz luminosity, and again there is a continuous variation in stellar mass and D$_\textrm{n}$(4000) with accretion rate.

The \emph{WISE} colour-colour diagram shows that the mid-infrared colours of a source vary with accretion rate. There is a clear trend for the sources with the lowest accretion rates to have the bluest W2-W3 colours, and therefore the oldest stellar population, supporting the D$_{\rm n}$4000 trend seen in Fig.~\ref{fig:propertiesL14_colour}, and for the sources with the highest accretion rates to have the reddest W1 - W2 colours, and therefore the largest contribution from the AGN. 
Accretion rate seems to be a better predictor of where a source will be found on this colour-colour diagram than HERG or LERG class (compare to Fig.~\ref{fig:WISEcolours}). This figure suggests that there might be a continuous distribution of mid-infrared source properties which vary with accretion rate, rather than a dichotomy.

%------------------------------------------------------------------------------%
\section{Discussion}\label{section:discussion}

\subsection{Accretion rates}

The results presented in Section~\ref{section:ratesHERGs} show that HERGs have higher Eddington-scaled accretion rates than LERGs, but there is some suggestion that there is more overlap between the two classes than has been found in previous studies (e.g.\ \citealt{2012MNRAS.421.1569B,2014MNRAS.440..269M,2015MNRAS.447.1184F}). Our work extends to lower radio luminosities than these studies, but we find some overlap between the populations at a range of luminosities. We see no evidence for two populations in the distributions of either [OIII] equivalent width or Excitation Index, the two main parameters used to identify HERGs and LERGs. This suggests that any division chosen in these parameters is perhaps arbitrary. This is in contrast to the study by \citet{2010A&A...509A...6B} at higher radio luminosities, who report a clear dichotomy in EI values. This points to a possible scenario where radio galaxies exhibit a broad range of accretion rates, and the stellar masses and stellar ages of these galaxies are correlated with this accretion rate. A possible mechanism for this correlation is the gas supply, as has been widely suggested in the literature, with galaxies which have an abundant supply of cold gas having a high accretion rate and a young stellar population, while sources with a limited supply of cold gas accrete much more slowly and have an older stellar population. We, however, find no compelling evidence for a `switch' between these two scenarios in our data and instead suggest a continuous distribution.

The rate at which galaxies accrete matter, and the mechanism by which they deposit energy back out again, has implications for models of galaxy formation. Understanding the different accretion modes, or lack of them, is clearly vital to characterising feedback accurately in these models.

\subsection{Implications for AGN feedback}

AGN feedback is a key component of simulations of galaxy formation as it is required to quench star-formation (e.g.\ \citealt{2006MNRAS.370..645B,2006MNRAS.365...11C}). Current hydrodynamical simulations (Horizon-AGN; \citealt{2014MNRAS.444.1453D}, Illustris; \citealt{2014MNRAS.444.1518V}, EAGLE; \citealt{2015MNRAS.446..521S}, MUFASA; \citealt{2016MNRAS.462.3265D}) all assume that the energy deposited back into the interstellar medium (ISM) scales directly with AGN accretion rate. Some simulations implement quasar and radio mode feedback separately with different efficiencies (e.g.\ Horizon-AGN, Illustris) while others do not (e.g.\ EAGLE, MUFASA).

\begin{figure}
\centerline{\includegraphics[width=\columnwidth]{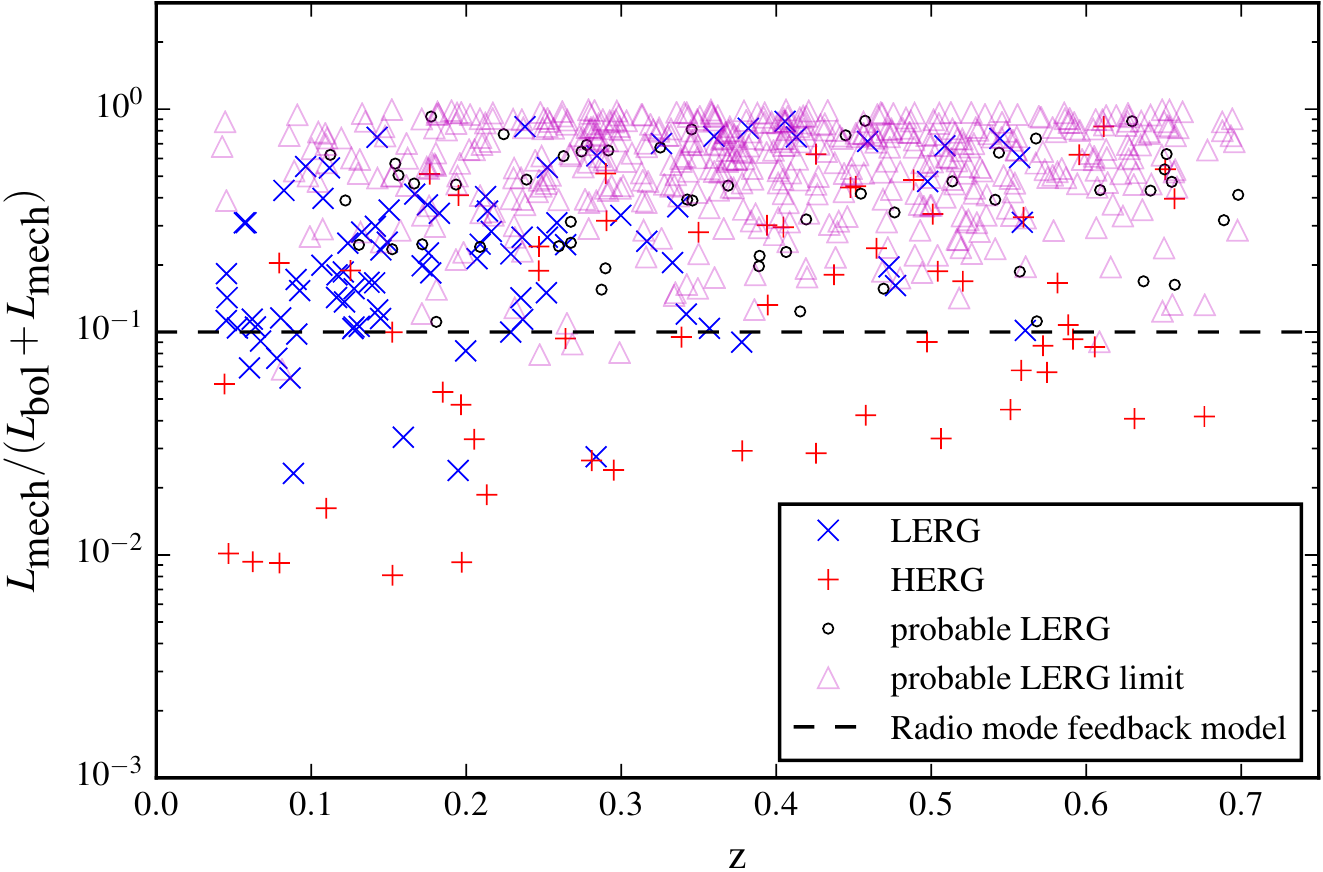}}
\caption{Fraction of the accreted energy released in the jets for the different source types. Triangles represent sources with an upper limit on their radiative accretion rate, so the fraction of energy released in the jet is a lower limit. The dashed line is the radio mode feedback model used in Horizon-AGN from \citet{2014MNRAS.444.1453D}. The uncertainties in the scaling relations used to estimate $L_\textrm{bol}$ and $L_\textrm{mech}$ are 0.4 and 0.7 dex respectively.
}\label{fig:Lmech}
\end{figure}

Fig.~\ref{fig:Lmech} shows $L_\textrm{mech} / (L_\textrm{bol} + L_\textrm{mech})$, which provides a measure of the fraction of the total accreted energy released back into the ISM in mechanical form in radio jets. As an example, in Horizon-AGN the mechanical energy deposition rate is modelled as $\epsilon_f \epsilon_r \dot{M}_\textrm{BH} c^2$, where $\epsilon_r = 0.1$ and $\epsilon_f = 1$, which is shown as the dashed line in Fig.~\ref{fig:Lmech}. It is clear from this Figure that for many of the sources the effect of mechanical feedback is significantly underestimated by this model. This is particularly true for LERGs, thought to be predominantly responsible for the maintenance mode of mechanical feedback at $z<1$, with 87 per cent of the LERGs and probable LERGs depositing more than 10 per cent of their accreted energy back into the ISM in mechanical form. This Figure also shows that radio-mode/mechanical feedback can be significant for HERGs as well as for LERGs; 28 out of 60 HERGs in this sample release more than 10 per cent of their accreted energy in their jets. However, the fact that these jets are generally more collimated my reduce the overall impact on their environment.

There is a scatter of $\sim 2$~dex in the fraction of the energy released in the jets, so the assumption that there is a direct scaling between accretion rate and jet power used in hydrodynamical simulations does not necessarily hold. This could be because the mechanical power output by an AGN is influenced by environment, and therefore does not depend only on accretion rate. \citet{2013ApJ...770..136I,2015MNRAS.453.2682I} found strong correlations between radio luminosity, cluster richness and central cluster density for LERGs, which supports this idea. The effect of environment for our sample will be investigated further in a future paper.

%------------------------------------------------------------------------------%
\section{Conclusions}\label{section:conclusions} 

We have used optical spectra to classify radio galaxies in the \citeauthor{2016MNRAS.460.4433H} 1 - 2~GHz VLA Stripe~82 Snapshot Survey as HERGs and LERGs. The radio data covers 100 square degrees and is a factor of two deeper than FIRST, allowing us to probe less powerful sources than studies based on FIRST or NVSS. This has resulted in a sample of 60 HERGs, 149 LERGs and 600 probable LERGs. Using optical spectra and mid-infrared data to investigate the properties of these HERGs and LERGs we find that at $z<0.3$ HERGs are hosted by galaxies with lower masses than LERGs. The HERGs in our sample are hosted by galaxies with a younger stellar population across the full range of redshift and luminosities probed here ($0.01 < z < 0.7$ and $10^{21} < L_{1.4~\rm GHz} / \textrm{ W \, Hz}^{-1} < 10^{27}$). This is consistent with previous results in the literature.

We find that HERGs tend to have higher Eddington-scaled accretion rates than LERGs, in agreement with previous results (e.g.\ \citealt{2012MNRAS.421.1569B,2014MNRAS.440..269M,2015MNRAS.447.1184F}). However, there is some suggestion of more overlap between the accretion rates of the two classes in our sample. We find that the host galaxy properties vary continuously with accretion rate, with the most slowly accreting sources being hosted by the most massive galaxies with the oldest stellar population, while the sources accreting matter more rapidly have lower-mass hosts and younger stellar populations. This is consistent with the idea that the accretion rate of a galaxy is linked to the supply of cold gas, with the sources with a readily available gas supply having high accretion rates and young stellar populations.

We find that the efficiency with which energy is deposited back into the ISM in radio jets is underestimated in radio AGN feedback models in many hydrodynamical simulations. A common assumption is that this efficiency is 10 per cent of the accretion rate; we find that 84 per cent of sources have an efficiency greater than this. We also find that radio mode feedback is significant for HERGs as well as for LERGs; nearly half of the HERGs in our sample release more than 10 per cent of their accreted energy in their jets. There is a scatter of $\sim 2$~dex in the fraction of accreted energy released in mechanical form so the assumption used in most hydrodynamical simulations that AGN feedback scales directly with accretion rate does not necessarily hold.

%------------------------------------------------------------------------------%
\section*{Acknowledgements}

We thank the anonymous referee for their contribution to this paper. IHW, MP, KM and MJJ acknowledge the financial assistance of the South African SKA Project (SKA SA) towards this research (www.ska.ac.za). IHW thanks the South African Astronomical Observatory, where some of this work was carried out. This research has made use of NASA's Astrophysics Data System. This publication makes use of data products from the \emph{Wide-field Infrared Survey Explorer}, which is a joint project of the University of California, Los Angeles, and the Jet Propulsion Laboratory/California Institute of Technology, funded by the National Aeronautics and Space Administration. This research made use of Astropy, a community-developed core Python package for Astronomy \citep{2013A&A...558A..33A}.
%IHW and MJJ acknowledge support from the Square Kilometre Array South Africa. %
Funding for SDSS-III has been provided by the Alfred P. Sloan Foundation, the Participating Institutions, the National Science Foundation, and the U.S. Department of Energy Office of Science. The SDSS-III web site is http://www.sdss3.org/.
%SDSS-III is managed by the Astrophysical Research Consortium for the Participating Institutions of the SDSS-III Collaboration including the University of Arizona, the Brazilian Participation Group, Brookhaven National Laboratory, University of Cambridge, Carnegie Mellon University, University of Florida, the French Participation Group, the German Participation Group, Harvard University, the Instituto de Astrofisica de Canarias, the Michigan State/Notre Dame/JINA Participation Group, Johns Hopkins University, Lawrence Berkeley National Laboratory, Max Planck Institute for Astrophysics, Max Planck Institute for Extraterrestrial Physics, New Mexico State University, New York University, Ohio State University, Pennsylvania State University, University of Portsmouth, Princeton University, the Spanish Participation Group, University of Tokyo, University of Utah, Vanderbilt University, University of Virginia, University of Washington, and Yale University.

%------------------------------------------------------------------------------%
%
% using custom format in ADS:
%
% %z132 \\bibitem[%\2m%(y)%\3m]%{R}\n   %\8.1g,%\Y,%\q,%\V,%\p

\setlength{\labelwidth}{0pt}

\bsp

\label{lastpage}
%------------------------------------------------------------------------------%
\end{document}